\documentclass[onecolumn,floats,showpacs,prd,amssymb,floatfix,nofootinbib,balancelastpage,superscriptaddress,amsmath,showkeys]{revtex4}

\pdfoutput=1

\usepackage{epsfig}
\usepackage{latexsym}
\usepackage{graphicx}
\usepackage{amssymb}
\usepackage{subfigure}
\usepackage{hyperref}
\def\sq{\displaystyle{\not} q} 
\def\sp{\displaystyle{\not} p} 
\def\sk{\displaystyle{\not} k}

\def\sv{\displaystyle{\not} v}

\def\beqn{\begin{eqnarray}} 
\def\eeqn{\end{eqnarray}} 
\def\be{\begin{equation}}
\def\ee{\end{equation}}
\def\nn{\nonumber}

\begin{document}
 \DeclareGraphicsExtensions{.pdf,.png,.gif,.jpg}

\title{Probing Dark Matter with AGN Jets}

\author{Mikhail Gorchtein}
\email{mgorshte@indiana.edu}
\affiliation{%
Center for the Exploration of Energy and Matter, Indiana University, Bloomington, IN 47408
}

\author{Stefano Profumo}
\email{profumo@scipp.ucsc.edu}
\affiliation{%
Santa Cruz Institute for Particle Physics and Department of Physics,\\ University of California, Santa Cruz CA 95064
}

\author{Lorenzo Ubaldi}
\email{ubaldi@physics.ucsc.edu}
\affiliation{%
Santa Cruz Institute for Particle Physics and Department of Physics,\\ University of California, Santa Cruz CA 95064
}


\begin{abstract}
\noindent We study the possibility of detecting a signature of particle dark matter in the spectrum of gamma-ray photons from active galactic nuclei (AGNs) resulting from the scattering of high-energy particles in the AGN jet off of dark matter particles. We consider particle dark matter models in the context of both supersymmetry and universal extra-dimensions (UED), and we present the complete lowest-order calculation for processes where a photon is emitted in dark matter-electron and/or dark matter-proton scattering, where electrons and protons belong to the AGN jet. We find that the process is dominated by a resonance whose energy is dictated by the particle spectrum in the dark matter sector (neutralino and selectron for the case of supersymmetry, Kaluza-Klein photon and electron for UED). The resulting gamma-ray spectrum exhibits a very characteristic spectral feature, consisting of a sharp break to a hard power-law behavior. Although the normalization of the gamma-ray flux depends strongly on assumptions on both the AGN jet geometry, composition and particle spectrum as well as on the particle dark matter model and density distribution, we show that for realistic parameters choices, and for two prominent nearby AGNs (Centaurus A and M87), the detection of this effect is in principle possible. Finally, we compare our predictions and results with recent gamma-ray observations from the Fermi, H.E.S.S. and VERITAS telescopes. 
\end{abstract}


\maketitle

\section{Introduction}\label{sec:intro}

The innermost regions of active galactic nuclei (AGNs) correspond to locations where the highest dark matter densities in the universe are believed to occur. This remarkable feat, and the fact that AGNs are well-known sources of high-energy particles, naturally leads to the question of whether anything could happen and be observed as those high-energy particles in the AGN jets traverse high-density dark matter regions. One possibility, first cleverly suggested some time ago by Bloom and Wells in Ref.~\cite{Bloom:1997vm}, is that, for AGN jets pointing off of our line-of-sight, an isotropic photon emission might result from the scattering of the high energy electrons in the jet off of dark matter particles. The estimates of Ref.~\cite{Bloom:1997vm}, unfortunately, and for reasons we will discuss in the present analysis, led to rather pessimistic conclusions, and this novel route to search for particle dark matter was basically dismissed.

With the advent of the Fermi Large Area Telescope (LAT) \cite{fermi}, and with the amazing recent results from ground-based Cherenkov Telescopes, most notably H.E.S.S. \cite{Aharonian:2009xn}, we deem it timely to re-consider the original proposal of Bloom and Wells. In particular, we here carry out a more in-depth analysis and we improve on the results of Ref.~\cite{Bloom:1997vm} in several respects, including:
\begin{enumerate}
\item a complete calculation of the relevant cross section for neutralinos in the minimal supersymmetric extension of the Standard Model and for the lightest Kaluza-Klein particle of Universal Extra-Dimensions
\item data-driven and semi-analytic models for the dark matter density distribution of the target AGNs
\item observationally motivated jet models
\item a novel discussion on proton jets and proton-dark matter scattering
\item a comparison with the recent Fermi-LAT and H.E.S.S. data.
\end{enumerate}
Perhaps more crucially than the refinements listed above, though, the main novel theoretical fact we point out here is that the electron (or, as we shall also consider here, proton) scattering off of dark matter produces a unique spectral feature, physically associated to a resonance that is present for several motivated particle dark matter models. This property of the cross section was not recognized in the seminal work of Ref.~\cite{Bloom:1997vm} due to a simplifying assumption in the basically only dimensional estimate of the relevant cross section. The resonant channels we find, quite interestingly, dominate the overall size of the scattering cross section, yielding much larger photons fluxes than those estimated in Ref.~\cite{Bloom:1997vm}. 

Observationally, the predicted spectral feature corresponds to a sharp cut-off in the spectrum of the outgoing photons, a sudden spectral break that might in principle be detectable by both Fermi and Cherenkov Telescopes observations. The energy corresponding to the cut-off depends non-trivially on the particle spectrum of the dark sector, and specifically on two key mass scales, including the mass of the dark matter particle. Whether or not this predicted signature will actually manifest itself in gamma-ray observations depends however critically on the details of the AGN jets and on the dark matter density distribution, in addition to the dark matter particle model. Also, disentangling this feature from off-jet astrophysical gamma radiation in the AGN is a highly non-trivial exercise, which we leave to future studies, whereas the focus of the present study is limited to the structure and intensity of the exotic dark matter related signal.

The differential photon flux from electron or proton (e[p]) scattering off of dark matter particles is simply expressed as the following integral over the energy of the impinging e[p]'s:
\be \label{eq:flux}
\frac{d\Phi_\gamma}{dE_\gamma}=\int  \left(\delta_{\rm
  DM} \right)\times \left(\frac{1}{d^2_{\rm
      AGN}}\frac{d\Phi^{\rm AGN}_{e[p]}}{dE_{e[p]}}\right)\times \left(\frac{1}{M_\chi}
  \frac{d^2\sigma_{e[p]+\chi \rightarrow \gamma + \dots}}{d\Omega
    dE_\gamma}\right)_{\cos \theta_0}\ {\rm d}E_{e[p]}.
\ee
The integrand is the product of three factors: 
\begin{enumerate}
\item the first factor, $\delta_{\rm DM}$, involves the dark matter density profile in the AGN, and is defined in Eq.~(\ref{eq:profile}) below;
\item the second factor depends upon the target AGN, and in particular on the AGN distance $d_{\rm AGN}$ and on the spectrum of the electron or proton flux;
\item the third factor, finally, depends
upon the dark matter particle model, and it involves the total cross section for
e[p]-$\chi$ scattering (where $\chi$ indicates the dark matter particle) into any final state that contains a photon, divided
by the dark matter particle mass $M_\chi$ (the cross section is computed at a
scattering angle $\theta_0$ between the direction of the AGN jet and the
line of sight).
\end{enumerate}
Conceptually, our analysis addresses, one after the other, the three factors listed above, and eventually leads to estimates of the full integral in Eq.~(\ref{eq:flux}). For the sake of comparing theoretical estimates with actual gamma-ray observations, choices need to be made for both the target AGNs and the particle dark matter models. For the first, we focused on two nearby AGNs, M87 and Centaurus A, for which recent Fermi-LAT data were presented in Ref.~\cite{fermim87,fermicena}. Observations at other wavelengths for these two well-studied objects also allow for educated estimates for the relevant parameters for the AGN jets entering the second factor above. As far as dark matter particle models are concerned, we rely on two widely studied weakly interacting massive particle models: supersymmetric neutralinos \cite{susyrev} and the Kaluza-Klein $B^{(1)}$ particle from the Universal Extra Dimensional scenario \cite{uedrev}. On a side note, we remind the Reader that other dark matter particles, specifically axion-like particles, can also produce observable effects in the gamma-ray flux from AGNs, for instance via the attenuation produced by photon to axion-like particle conversion in the presence of a magnetic field near the source, as pointed out in the recent studies listed in Ref.~\cite{alp}.

The organization of the paper is as follows: In section~\ref{sec:DMprofile} we discuss the estimate of the dark matter density profile and the calculation of the first factor in Eq.~(\ref{eq:flux}); In
section~\ref{sec:ed} we discuss the spectral distribution of electrons and protons in the AGN jet, that enter the
second factor of the above integral. We then calculate and study the cross
section for the process of interest in
section~\ref{sec:cross}, that will allow us to calculate the third and last factor of the integrand of Eq.~(\ref{eq:flux}). Finally, we put all three factors together to
compute the total photon flux in section~\ref{sec:flux}, and we discuss
our results in relation to the sensitivity of current telescopes, and we especially compare our results with the recent Fermi-LAT observations \cite{fermim87,fermicena}. We devote sec.~\ref{sec:protons} to a discussion of the case of proton jets, and we conclude in sec.~\ref{sec:conclusions}. We leave some considerations on relativistic effects in the particle energy distribution in the jet to Appendix A and the details of the cross section calculation to Appendix B.

\newpage

\section{The First Factor: Dark matter density profile} \label{sec:DMprofile}
The first factor in Eq.~(\ref{eq:flux}) corresponds to the line-of-sight integral of  the dark matter density in the AGN under consideration, i.e. to an effective average of the dark matter density times a relevant length-scale associated to the AGN jet:
\begin{equation} \label{eq:profile}
\delta_{\rm DM} \equiv <\rho_{\rm DM}R_{\rm DM}> = \int_{r_{min}}^{r_0} \rho_{\rm DM}(r)\  {\rm d}r,
\end{equation}
where the integration runs between $r_{min}$,  the minimum distance from the AGN center at which the scattering process we study takes place (thus, effectively, the base of the AGN jet), and $r_0$, an upper limit of integration that corresponds to the distance at which the AGN jet peters out. Also, in the Equation above, $\rho_{\rm DM}(r)$ is the dark matter density profile as a function of the distance from the central black hole. We expect only a mild dependence on the precise value of the upper limit of integration $r_0$, since dark matter density profiles fall off steeply with radius, whereas it will be important for our calculation to model as accurately as possible the innermost portion of the dark matter distribution, taking into account the gravitational effects of the central super-massive black hole on the equilibrium density distribution. 

The two AGNs we consider in our study, Centaurus A and M87, lie at a estimated
 distances of 3.7 and 16 Mpc, respectively \cite{Romanowsky:2000zb, Kraft:2003gp}. Such distances are large enough
to make it impossible to have observations that can resolve regions as close as 100 or 1000 Schwarzschild radii from the compact object in the AGN core. In addition, dynamical tracers of the dark matter density distribution also usually provide very limited information at small radii. We thus necessarily need to rely on extrapolations to the dark matter density distribution, which we base on both results of N-body simulations and on theoretical studies on the effect on the dark matter density profile of the presence of a compact object (the AGN super-massive black hole) and of the surrounding baryons and stars. To this end, we adopt the theoretical results presented in Ref.~\cite{Gondolo:1999ef} and \cite{Gnedin:2003rj}. 

Working under the assumptions that the dark
matter particles are collisionless and that the central black hole
grew adiabatically by accretion of gas and stars, Gondolo and Silk
\cite{Gondolo:1999ef} found that the
dark matter forms a dense central spike. The profile they obtain for an initial dark matter density distribution with a power-law inner profile $\rho_{\rm ini}(r)\propto r^{-\gamma}$ reads:
\begin{equation} \label{eq:GondSilk}
\rho_{\rm DM}(r) = \rho_{\rm sp}(r) = \frac{\rho'(r) \rho_{\rm
    core}}{\rho'(r) + \rho_{\rm core}},
\end{equation}
where
\begin{equation}
\rho_{\rm core} \simeq M_\chi / (\langle\sigma v\rangle_0 t_{BH}) \qquad \qquad
\rho'(r) = \rho_R g_\gamma(r) \left( \frac{R_{\rm sp}}{r}
\right)^{\gamma_{\rm sp}}.
\end{equation}
In the Equation above, $\rho_R = \rho_0 \left( \frac{R_{\rm sp}}{r_0}
\right)^{-\gamma}$, $g_\gamma(r) \simeq \left( 1 - \frac{4 R_S}{r}
\right)^3$, $R_{\rm sp} = \alpha_\gamma r_0 \left(
  \frac{M_{BH}}{\rho_0 r_0^3} \right)^{\frac{1}{3-\gamma}}$,
$\gamma_{\rm sp} = \frac{9-2\gamma}{4-\gamma}$, with $\alpha_\gamma$ a numerical coefficient that depends on the initial dark matter density profile slope $\gamma$. $\rho_{\rm DM}$ vanishes for $r<4R_S$, particles on smaller orbits being accreted on the super-massive black hole \cite{Gondolo:1999ef}. $r=4R_S$ thus sets the lowest possible value for $r_{min}$.
Some numerical values for quantities that enter the profiles defined above are listed in
Table~\ref{tab:profpar} for Centaurus A and M87, together with some nominal assumptions on the particle dark matter models. In particular,  in Tab.~\ref{tab:profpar} we consider annihilation rates with an upper bound set by the typical pair-annihilation cross section required by obtaining a thermal relic abundance close to the cosmological dark matter density, $\langle\sigma v\rangle_0\sim10^{-26}$ cm$^{-3}$s$^{-1}$, and with a lower bound close to what found in the case of bino-like neutralinos in the MSSM with co-annihilations setting the required relic density to the desired value \cite{Profumo:2004at}, $\langle\sigma v\rangle_0\sim10^{-30}$ cm$^{-3}$s$^{-1}$.

\begin{table}[!b]
\begin{ruledtabular}
\begin{tabular}{l r r}
 {} &  \qquad Centaurus A & \qquad \qquad M87 \\
\hline \hline
$M_\chi$ [GeV]  DM mass & $100$ & $100$  \\
$M_{\rm BH} \ [{\rm M}_\odot]$  Black Hole mass & $(5.5\pm 3.0) \times
10^7$ & $(6.4\pm 0.5)\times 10^9$ \\
$R_S$ [pc]  Schwarzschild radius & $5\times 10^{-6}$ & $6\times 10^{-4}$ \\
$t_{\rm BH}$ [yr]  Black Hole time & $10^8 \div 10^{10}$  & $10^8 \div 10^{10}$  \\
$\langle\sigma v\rangle_0$ [cm$^3/$s]   Annihilation cross section& $10^{-30} \div 10^{-26}$  & $10^{-30} \div
10^{-26}$ \\
 $\alpha_\gamma$ & 0.1 & 0.1 \\
$r_0$ [kpc]  upper limit of integration & 15 & 15 
\end{tabular}
\caption{Values for parameters relevant to the Gondolo \& Silk dark matter profile. The black
  hole mass is from Ref.~\cite{Neumayer:2010kw} for Centaurus A, and
  from Ref.~\cite{Gebhardt:2009cr} for M87. \label{tab:profpar}}
\end{ruledtabular}
\end{table}

A few years after the work of Gondolo and Silk, Gnedin and Primack
\cite{Gnedin:2003rj} pointed out that the central
regions around super-massive black holes usually harbor stars that can gravitationally scatter the dark matter particles and
cause the distribution to evolve towards an equilibrium
solution that differs from what found in Ref.~\cite{Gondolo:1999ef}. Taking scattering off of stars into account, Gnedin and Primack found a universal
inner profile that scales as $r^{-3/2}$, which we write as
\be \label{eq:joel}
\rho_{\rm DM}(r) = \rho_0 \left(\frac{a}{r}\right)^{3/2},
\ee
where we set $a=10^5 \ R_S$. Notice that the Gnedin and Primack profile of Eq.~(\ref{eq:joel}) is a more conservative choice than the one taken in Ref.~\cite{Bloom:1997vm}, where $\rho_{\rm DM}(r)\sim r^{-1.8}$.

The profiles of Equations (\ref{eq:GondSilk}) and (\ref{eq:joel}) are
assumed to be valid only in the innermost AGN regions, which are also those most relevant to the calculation of the quantity in Eq.~(\ref{eq:profile}). To normalize the profiles, we need an estimate for the enclosed dark matter mass in the innermost regions around the central point sources. At radii relevant for the determination of the black hole masses, typically $10^5 \ R_S$  \cite{Neumayer:2010kw, Gebhardt:2009cr}, we require that the enclosed dark matter mass be, at most, as large as the uncertainty over the black hole mass, i.e.: 
\be \label{eq:MenCenA}
 \int_{r_{low}}^{10^5 R_S} dr 4\pi r^2
\rho_{\rm DM} \leq 3\times 10^7 \ {\rm M}_\odot
\ee
 for Centaurus A and
\be \label{eq:MenM87}
\int_{r_{low}}^{10^5 R_S} dr 4\pi r^2
\rho_{\rm DM} \leq 5\times 10^8 \ {\rm M}_\odot
\ee
 for M87. Note that
the lower limit of integration, $r_{low}=4R_S$ is obvious for the profile of
Eq.~(\ref{eq:GondSilk}), while for  the
profile of Eq.~(\ref{eq:joel}) the lower cutoff will be set by the requirement that the dark matter not annihilate away during the black hole lifetime, which corresponds to densities $\rho \simeq M_\chi / (\langle\sigma v\rangle_0 t_{BH})$. For the AGN and particle dark matter parameters we consider, $r_{low}=100 R_S$ is an appropriate choice for the profile by Primack and Gnedin. 

Using the Gondolo and Silk
distribution, we notice that one runs in the additional complication that the integrands
in (\ref{eq:MenCenA}) and (\ref{eq:MenM87}) depend non-linearly on
$\rho_0$, when determining the latter quantity from Eq.~(\ref{eq:MenCenA}-\ref{eq:MenM87}) above. We note however that the main
contribution to the integral comes from $r \gg R_S$. In that regime we
have $\rho_{\rm core} \gg \rho'(r)$ so we can approximate $\rho_{\rm
  DM}(r) \simeq \rho'(r)$. We can then rewrite $\rho'(r) =
\rho_0^{\frac{1}{4-\gamma}} \left( \frac{R'_{\rm
      sp}}{r_0}\right)^{-\gamma} g_\gamma(r) \left( \frac{R'_{\rm
      sp}}{r} \right)^{\gamma_{\rm sp}}$, where $R'_{\rm sp} =
\alpha_\gamma r_0 \left(  \frac{M_{BH}}{r_0^3}
\right)^{\frac{1}{3-\gamma}}$. Now Equations (\ref{eq:MenCenA}) and (\ref{eq:MenM87}) are linear in
$\rho_0$ and can thus be easily solved. 

We note that the normalization used here is also consistent to available observational data. As a cross-check, we integrated our profiles to calculate the dark matter mass enclosed in 1 kpc and compared it to the mass profiles at small radii obtained from astronomical observations and given in Ref.~\cite{Kraft:2003gp} and Ref.~\cite{Romanowsky:2000zb} for Centaurus A and M87, respectively. In both cases we found the enclosed dark matter mass to be one to two orders of magnitude below the total mass at a radius of $\sim 1$ kpc, thus concluding that our normalization is reasonable and compatible with available data.

\begin{figure}[!t]
\centering
\subfigure[]{
\includegraphics[height=100 mm]{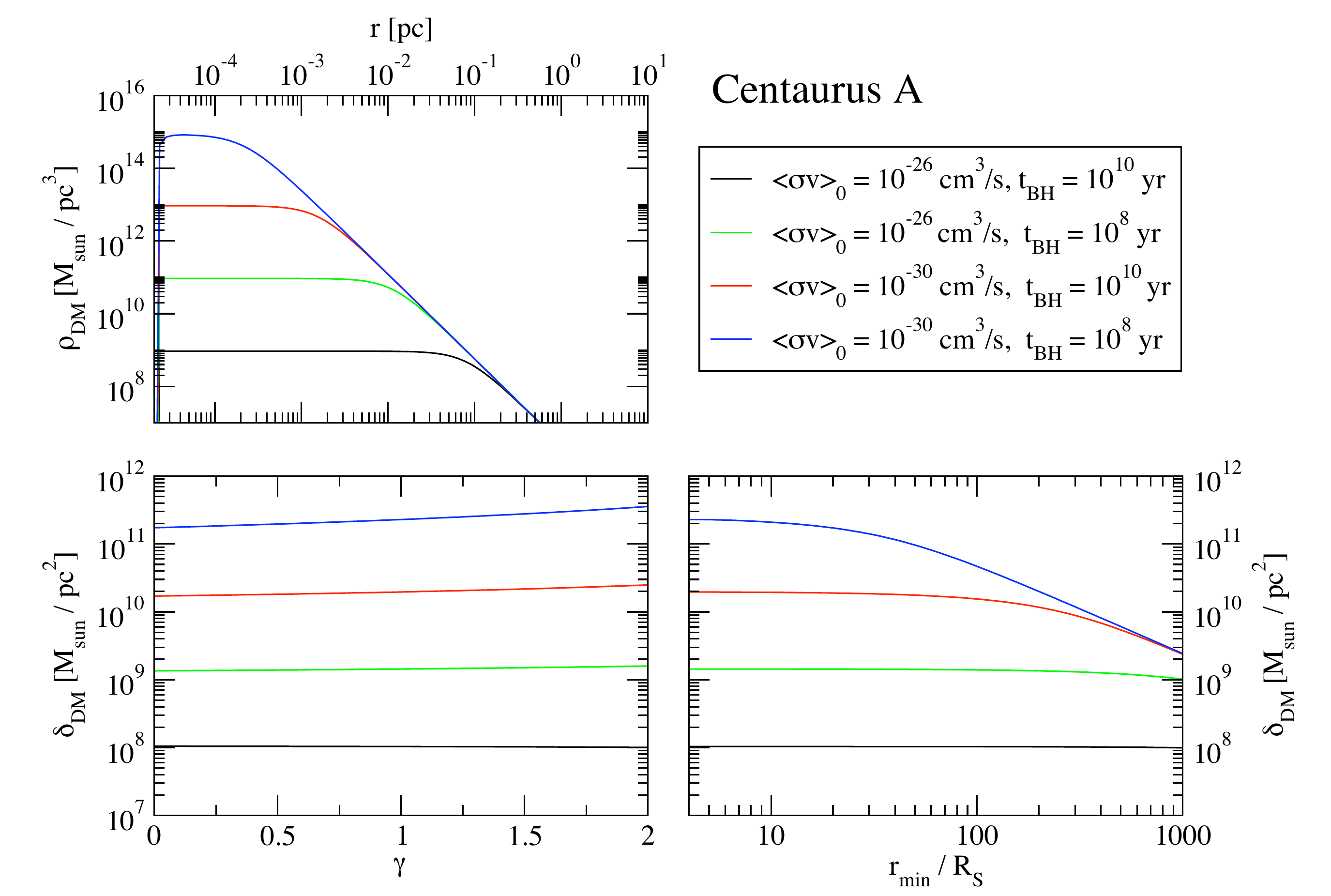} \label{fig:Cprofiles}}
\subfigure[]{
\includegraphics[height=100 mm]{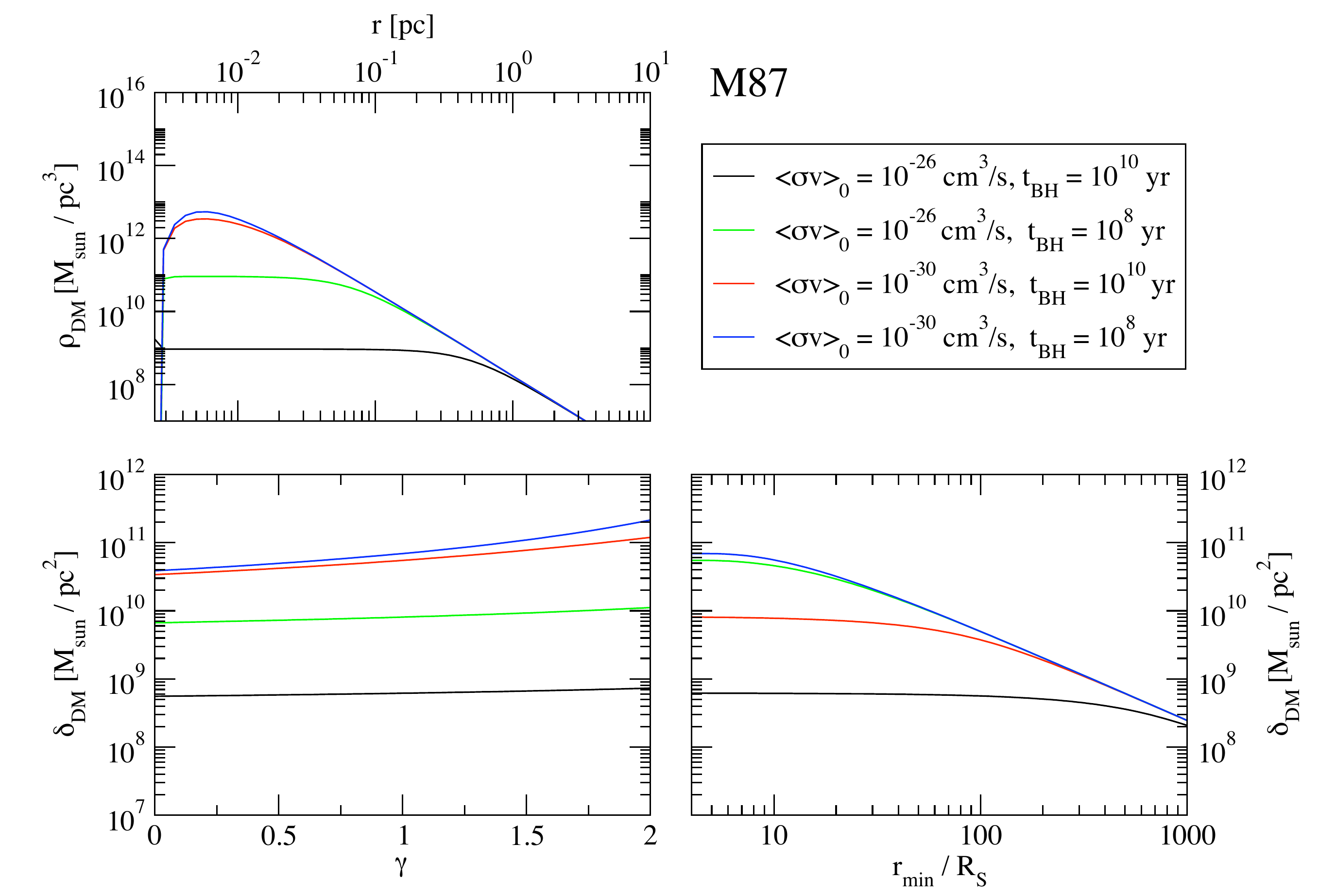} \label{fig:Mprofiles}}
\caption{{\it Top}. The dark matter density profile near the central AGN regions, for $\gamma=1$ and for four different choices for the parameters  $<\sigma v>_0$ and $t_{BH}$. {\it Bottom}. We show how the integral $\delta_{\rm DM}$
depends on the slope of the density profile prior to adiabatic compression $\gamma$ (left panel), and on the lower
limit of integration $r_{min}$, in units of $R_S$ (right panel). Figure (a) refers to Centaurus A, while (b) to M87.}
\end{figure}

We now are in a position to compare the values of the integral $\delta_{\rm DM}$ in Eq.~(\ref{eq:profile})
 - the first factor in the overall flux calculation outlined in Eq.~(\ref{eq:flux}) - corresponding to the two different profiles discussed. We show our results in Fig.~ \ref{fig:Cprofiles} and
\ref{fig:Mprofiles}. In each one of the two Figures, corresponding to Cen A and to M87, respectively, we show three panels: in the upper left panel we plot the dark matter density as a function of the distance from the AGN center. We set the original slope (prior to adiabatic compression) of the dark matter density profile $\gamma=1$. Notice that $\rho'(r)$ is the same
for all the curves shown. Also notice how the spike at small $r$ only emerges when the parameters are such that
$\rho_{\rm core} \gg \rho'(r)$, in which case $\rho_{\rm DM}(r) \simeq \rho'(r)$. The lower-left panel shows the (very mild) dependence of the $\delta_{\rm DM}$ factor on $\gamma$. Finally, the bottom-right panel illustrates how $\delta_{\rm DM}$ depends upon the assumed value of the distance between the origin of the AGN jet and the AGN central region, in units of $R_S$. Again, we set $\gamma=1$, which, incidentally, corresponds to $\gamma_{\rm sp}= 7/3$.

The four lines in Fig.~\ref{fig:Cprofiles} and \ref{fig:Mprofiles} correspond to different assumptions on the pair-annihilation rate of the dark matter particle (which we set to the two representative values $\langle\sigma v\rangle_0\sim10^{-26}$ cm$^{-3}$s$^{-1}$ and $\langle\sigma v\rangle_0\sim10^{-30}$ cm$^{-3}$s$^{-1}$ quoted and motivated above) and on the black hole age $t_{\rm BH}$ (for which we consider typical choices for the ages of $10^8$ yr and of $10^{10}$ yr). The combination $(\langle\sigma v\rangle_0 t_{BH})$ sets, for a given dark matter particle mass, the maximal density that survives complete annihilation, as can be seen directly from inspection of the upper-left panel. While the four lines all correspond to the density profiles described by Gondolo and Silk, where the density
of dark matter very close to the core is dominated by $\rho_{\rm core}$, resorting to
the profile by Gnedin and Primack we get values very close to the
lowest curves showed in Figures \ref{fig:Cprofiles} and
\ref{fig:Mprofiles}. Notice that a profile such as that employed in Ref.~\cite{Bloom:1997vm} would have fallen in between the lines we show in those figures.

We notice that the most promising dark matter models are those with a low pair-annihilation cross section: this is a quite remarkable feat, as this is the exact opposite of what is found for any other indirect dark matter detection method! Here, the reason is that the less the dark matter pair-annihilates, the more particles can be densely concentrated around the central regions of the AGN, and the more probable the scattering of electrons and protons off of these particles is. Also, the younger the central black hole, the better, since fewer particles have annihilated away since the black hole birth. Interestingly, the dependence on the distance of the base of the jet from the center of the AGN is relatively mild: only for extremely spiky profiles, such as the one represented by the blue line, do we find an order-of-magnitude dependence on $r_{\rm min}$, if the latter is within a few hundreds $R_S$, as is usually expected. Also, the slope of the dark matter density profile before adiabatic compression onto the black hole is virtually irrelevant, up to a factor 2 (lower-left panel). We also point out that the final value of the relevant parameter $\delta_{\rm DM}$ is mainly dictated by the innermost regions of the dark matter density profile. 

Comparing Cen A (Fig.~\ref{fig:Cprofiles}) with M87 (\ref{fig:Mprofiles}), we find that for the latter, featuring a significantly more massive central super-massive black hole, and hence a much larger $R_S$, the dependence on $r_{\rm min}$ is somewhat stiffer. Also, as apparent from the upper-left panel, in some cases the spike is cut off by the $r>4R_S$ limit beyond which particles infall in the black hole \cite{Gondolo:1999ef}. Overall, for the same particle physics model, a reasonable range of expected values for $\delta_{\rm DM}$ for Centaurus A falls between $10^8$ and $10^{11}$ in units of solar masses per pc$^2$, while a similar but slightly narrower range is expected for M87 ($3\times10^8\lesssim\delta_{\rm DM}/(M_\odot/{\rm pc}^2)\lesssim10^{11}$).

\section{The Second Factor: High Energy particles in the jet}\label{sec:ed}

The second factor in Eq.~(\ref{eq:flux}) depends upon the nature of the high-energy particle content of the AGN jet, including its composition, geometry, spectrum and overall energy content (``luminosity''). We assume that the jet consists of either or both relativistic electrons and
protons. For our purposes, the jet geometry is largely irrelevant: as we noted above, the exact position of the base of the jet is not fundamental
for our purposes as long as it lies within $\sim 100 \ R_S$, which is
believed to be the case for the AGNs under consideration here \cite{Orellana:2009nz}. The reason is that, as we saw in the previous section, the
$\delta_{\rm DM}$ integral is not much affected by a variation of the
lower limit of integration $r_{min}$, as long as $r_{min} < 100 \
R_S$. Other details of the jet geometry, such as the opening angle, for
example, are also not crucial to our calculation. One exception is the inclination of the jet with respect to the line of sight, which we will discuss in the next section, and which enters the angle at which the cross section is calculated. The crucial information we are after is, rather,
the energy distributions of electrons and protons that enter the
second factor in Eq.~(\ref{eq:flux}).

Let us focus on the case of electrons first. We adopt here the commonly used ``blob
geometry'' \cite{blobgeom}: the electrons move isotropically in the blob frame with a
power law energy distribution, and the blob itself moves with respect
to the central black hole with a moderate bulk Lorentz factor
$\Gamma_B=(1-\beta_B^2)^{-1/2}$. For Centaurus A and M87, $\Gamma_B$ is estimated to be $\sim
3$, although this depends on assumptions on e.g. the detailed gamma-ray emission mechanism (see e.g. the discussion in Ref.~\cite{fermicena}). In this section, primed quantities refer to the blob rest frame,
while unprimed quantities refer to the black hole rest frame.  
Motivated by jet models employed to reproduce the results of gamma-ray observations \cite{fermim87,fermicena}, we use a broken power law energy distribution for the relativistic particles in the jet:
\beqn
\frac{d\Phi_e^{\rm AGN}}{d\gamma'}(\gamma')  = \frac{1}{2} k_e
\gamma'^{-s_1}\left[1+\left(\frac{\gamma'}{\gamma'_{\rm br}}\right)^{s_2-s_1}\right]^{-1} \qquad {\rm for}  \quad \gamma'_{\rm min}\leq \gamma' \leq \gamma'_{\rm max}, 
\eeqn
where $\gamma'=E'/m_e$, and the constant $k_e$ can be evaluated from
the kinetic power of the jet, $L_e$. First we need to boost this
distribution to the black hole frame (details of this calculation are presented in Appendix~\ref{app:special}). We find
\begin{equation}
\int_{-1}^1 d\mu' \int_{\gamma'_{\rm min}}^{\gamma'_{\rm max}}d\gamma' \frac{d\Phi_e^{\rm
    AGN}}{d\gamma'}(\gamma') =\int_{-1}^1 \frac{d\mu}{\Gamma_B(1-\beta_B
  \mu)} \int _{\gamma_{\rm min}}^{\gamma_{\rm max}} d\gamma
\frac{d\Phi_e^{\rm AGN}}{d\gamma}(\gamma\Gamma_B(1-\beta_B \mu)),
\end{equation}
where $\mu'\equiv\cos \theta'$, $\mu \equiv \cos \theta$, $\theta'$ is the
polar angle of the electrons in the blob frame, $\theta$ is the polar
angle in the black hole frame. The distribution on the left hand side
is isotropic, as we mentioned, and does not depend on $\mu'$, but the
boosted distribution, on the right hand side, depends non trivially on
$\mu$.
Now we can evaluate the constant $k_e$ using
\begin{equation}
L_e=\int_{-1}^1  \frac{d\mu}{\Gamma_B(1-\beta_B
  \mu)} \int_{\gamma_{\rm min}}^{\gamma_{\rm max}} m_e^2 d\gamma \ \gamma \frac{d\Phi_e^{\rm AGN}}{d\gamma}(\gamma\Gamma_B(1-\beta_B \mu)).
\end{equation}
The values of the parameters that
appear in the Equations above are determined observationally and are taken here from the analyses of Ref.~\cite{fermicena} for Cen A and of Ref.~\cite{fermim87} for M87. They are summarized in Table~\ref{tab:eedpar}.

Notice that for the case of Cen A, in Ref.~\cite{fermicena} Abdo {\em et al.} point out how it is not possible to fit the Fermi and the H.E.S.S. data simultaneously using a single-zone synchrotron/synchrotron self Compton (SSC) model. One possibility mentioned there is that the H.E.S.S. emission originates from a different part of the jet than that responsible for the gamma rays observed by LAT. Our study suggests that another possibility, electron-dark matter scattering, is in principle viable to explain at least part of the Fermi gamma-ray detection. Thus, when choosing the appropriate normalization, we are not strictly bound to using a set of parameters that gives the best fit under the assumption of SSC. Looking, as a guidance, at the four different sets of parameters provided in Ref.~\cite{fermicena}, we employ the one that gives the best fit to the H.E.S.S. observations (since, with hindsight, the latter cannot be fitted by the emission model we consider here), but we use a slightly higher value for the jet power. As stated in Ref.~\cite{fermicena}, the jet powers quoted there are very conservative and can be considered as lower limits: It is thus safe to explore even higher values (and we will later on) as long as they are below the Eddington luminosity which, for CenA, is $\sim 10^{46}$ erg/s.

\begin{table}[!b]
\begin{ruledtabular}
\begin{tabular}{l c r r r r r}
 Parameter & Symbol &  Cen A & M87 \\
\hline \hline
Low-Energy Electron spectral index & $s_1$  & 1.8 & 1.6 \\
High-Energy Electron spectral index & $s_2$ &  3.5 & 3.6 \\
Minimum electron Lorentz factor & $\gamma'_{\rm min}$ & $8\times 10^2$ & $8\times 10^2$  \\
Break electron Lorentz factor & $\gamma'_{\rm br}$ &$4\times 10^5$&$4\times 10^3$\\
Maximum electron Lorentz factor & $\gamma'_{\rm max}$ & $10^8$ & $10^7$\\
Jet power in electrons [erg s$^{-1}$] &$L_e$& $3\times 10^{43}$&$10^{44}$ \\
\end{tabular}
\caption{Values for parameters in the electron energy
  distribution. For M87 the values are taken from
  Ref.~\cite{fermim87}, while for CenA they are from Ref.~\cite{fermicena}. Notice that in Ref.~\cite{fermicena} four
  different sets of parameters are used for four different fits. The
  one set we show here is the one that produces a good fit to the H.E.S.S. data, but, as explained in the text, we use a higher value for the jet power.
  \cite{Aharonian:2009xn}.)
\label{tab:eedpar}}
\end{ruledtabular}
\end{table}

Interestingly, we find that for $\Gamma_B \sim 3$ most of the electrons
are seen to move in a narrow cone around the direction of the jet in
the black hole frame (see
Appendix~\ref{app:special} for further details). 
When we compute the photon flux from Eq.~(\ref{eq:flux}), we will use for the second factor
\be
\frac{1}{d^2_{\rm
      AGN}}\frac{d\Phi^{\rm AGN}_{e}}{dE_{e}} = \frac{1}{d^2_{\rm
      AGN}}\int_{0.9}^1  \frac{d\mu}{\Gamma_B(1-\beta_B
  \mu)} \frac{d\Phi_e^{\rm AGN}}{d\gamma}(\gamma\Gamma(1-\beta_B \mu)) .
\ee 
Restricting the region of integration over $\mu$ from 0.9 to 1, one loses roughly 20\% of the electrons, while this choice restricts the high-energy scatterers to a quite collimated jet. The angle of the emitted photon with respect to the jet can now be considered to have an uncertainty of around $\pm \arccos (0.9)$,  comparable to (or much smaller than) the astrophysical uncertainties on the angle that the jet has with the line of sight, and therefore not a major concern for the accuracy of our final conclusions.

\section{The third Factor: Cross section} \label{sec:cross}
In this section we compute the cross section for the process ``electron
+ dark matter $\rightarrow$ electron + dark matter + photon''
using two different dark matter particle models: (1) the Minimal Supersymmetric
Standard Model (MSSM), where the dark matter candidate is assumed to be the lightest
neutralino, $\chi$ \cite{susyrev}; (2) Universal Extra Dimensions (UED), where the
dark matter is the lightest Kaluza-Klein particle (LKP), assumed to be the $B^{(1)}$ (first Kaluza-Klein mode of the hypercharge gauge boson $B$) \cite{uedrev}.  The structure of the calculation is as follows: We first perform a
full calculation for the supersymmetric model. We then infer from that calculation which
terms give the dominant contributions to the cross section, and we apply that knowledge to the second model, UED, for which we only
compute the analogous important terms.

\subsection{Neutralino Dark Matter}
\begin{figure}[!t]
\centering
\subfigure[${\cal M}_1$]{\label{fig:SUSY1} \includegraphics[width=50
  mm]{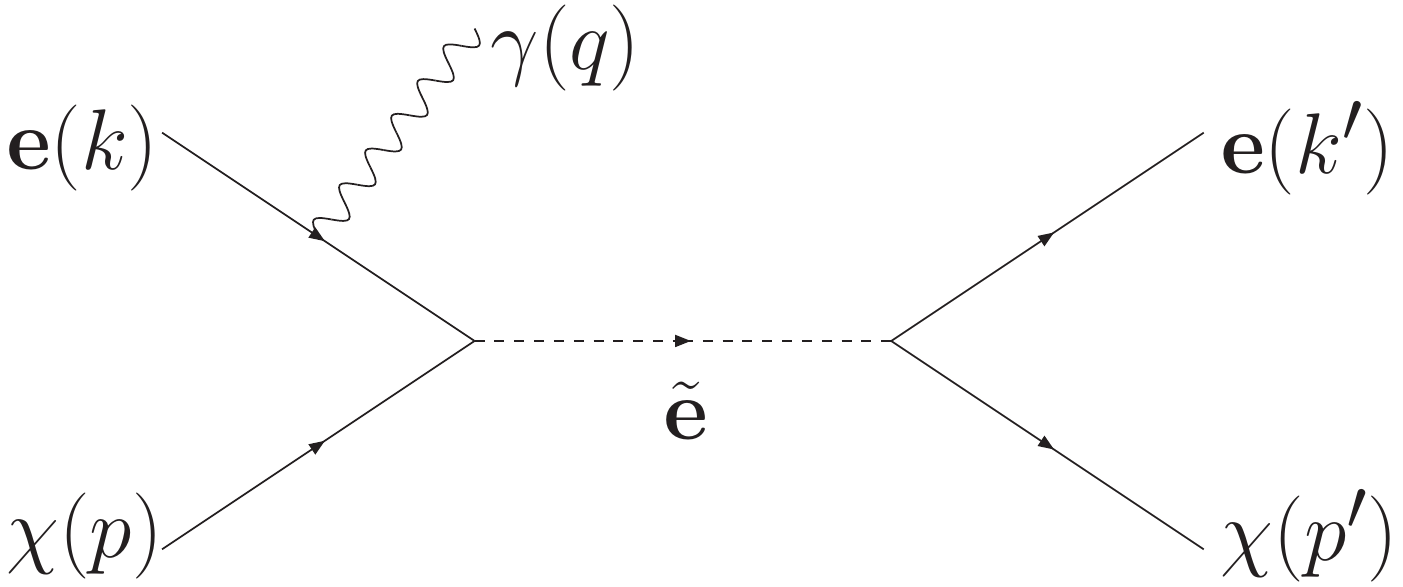} }
\hspace{5 mm}
\subfigure[${\cal M}_2$]{\label{fig:SUSY2} \includegraphics[width=50
  mm]{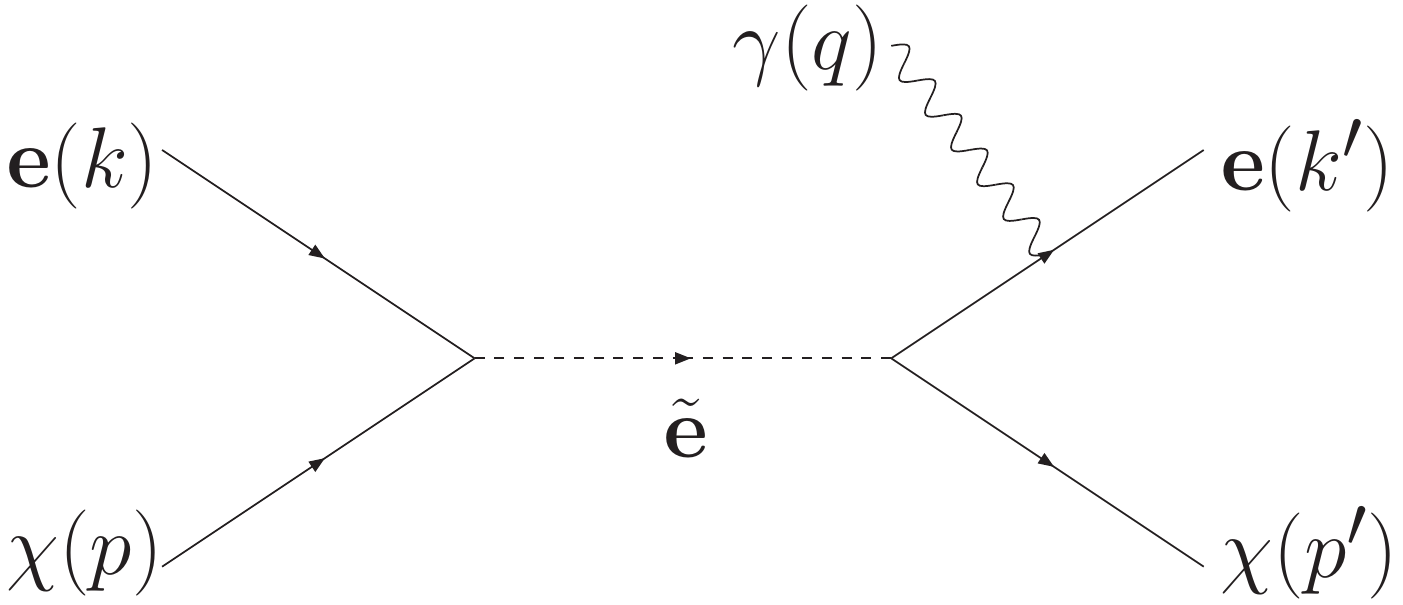}  }
\hspace{5 mm}
\subfigure[${\cal M}_3$]{\label{fig:SUSY3} \includegraphics[width=50
  mm]{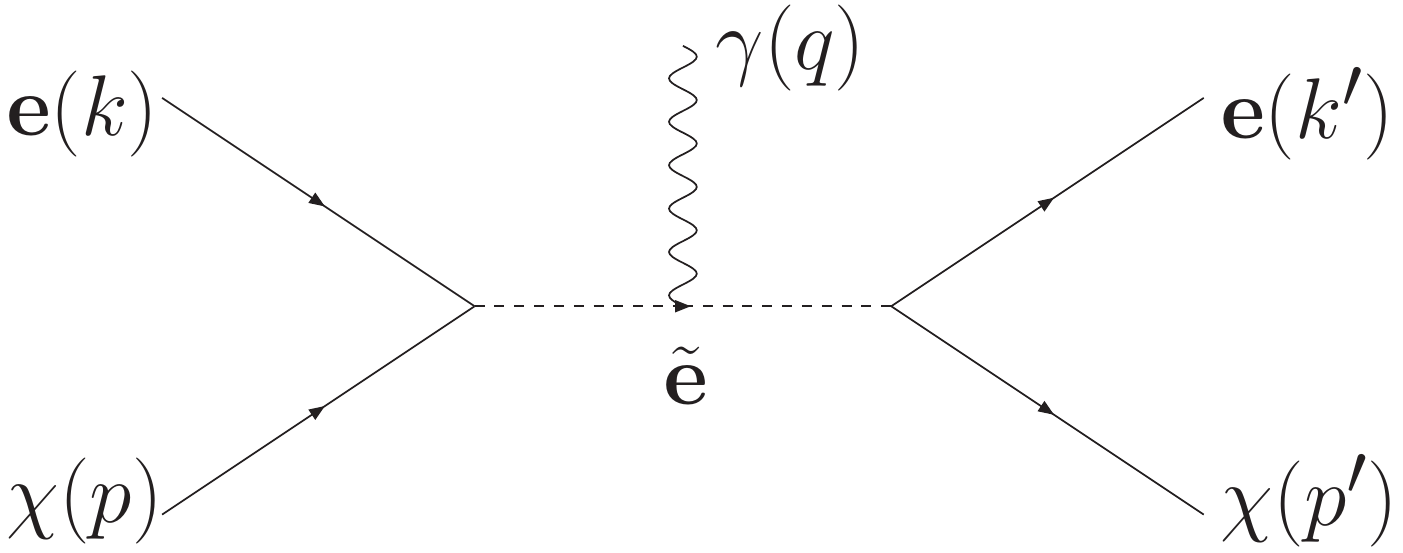} }
\caption{Feynman diagrams for the process $e(k)+\chi(p) \rightarrow e(k')+\chi(p')+\gamma(q)$.}
\label{fig:SUSYdiagrams}
\end{figure}

 We study the scattering process of electrons off of neutralinos with
 the emission of photons at the lowest significant order, i.e., symbolically, $e+\chi \rightarrow
 e+\chi+\gamma$. In principle there are several contributing Feynman diagrams, but without loss of generality and accuracy we can restrict our attention to the
 $s$-channel diagrams shown in Fig.~\ref{fig:SUSYdiagrams}. In fact, when
 the exchanged scalar electron (selectron) is on shell we have a resonance and these
 three diagrams are dominant, so it is fair to neglect the diagrams in the $t$- or $u$- channels, which never feature an on-shell particle in the intermediate state. Notice that no resonances at all appear from the two amplitudes considered in Ref.~\cite{Bloom:1997vm} - one of the crucial differences between the present results and the findings of that study. By only considering the process $e\chi\to \tilde e \gamma$, in fact, the intermediate state selectron can {\em never}, kinematically, be on-shell, and thus the process is never resonant.

Referring to the momenta as labeled in Fig.~\ref{fig:SUSYdiagrams}, we adopt the following notation:
\beqn
t&=&(k-q)^2\nn\\
t'&=&(k'+q)^2\nn\\
s&=&(p+k)^2\nn\\
s'&=&(p'+k')^2\nn\\
\Pi_s&=&\frac{1}{s-M_{\tilde e}^2-i\sqrt{s}\Gamma}\nn\\
\Pi_{s'}&=&\frac{1}{s'-M_{\tilde e}^2-i\sqrt{s'}\Gamma},
\eeqn
where $M_{\tilde e}$ is the selectron mass\footnote{For simplicity, we assume the right-handed and the left-handed selectrons to be degenerate in mass.} and
$\Gamma=\frac{1}{8\pi}(a^2_L+a^2_R)\frac{(M_{\tilde
    e}^2-M^2)^2}{4M_{\tilde e}^3}$ is the selectron energy-independent
width\footnote{See the appendix for the definition of the couplings
  $a_L$ and $a_R$.}, with $M_\chi$ the mass of the neutralino.  

\begin{figure}[!t]
\centering
\includegraphics[width=180mm]{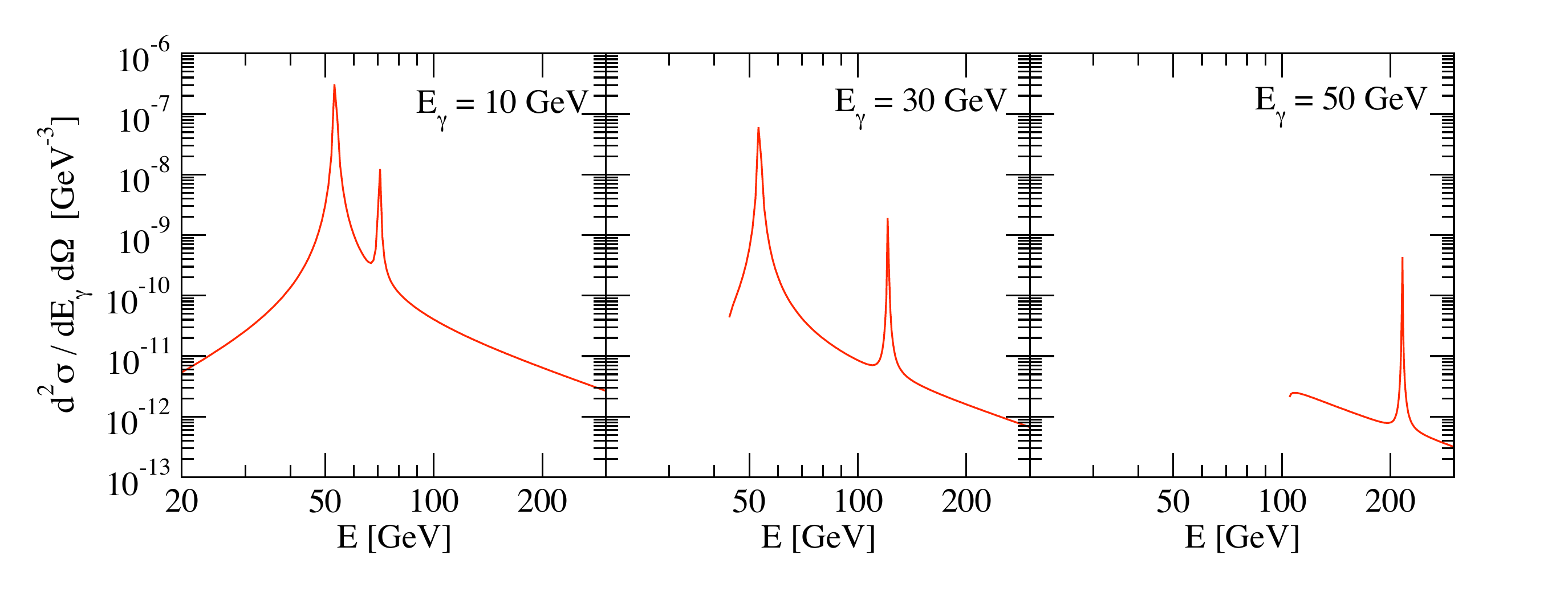} 
\caption{The differential cross section of
  Eq.~(\ref{eq:fullcross}) as a function of the initial electron energy
  $E$ for different values of the final photon energy $E_\gamma$. Note
  that we use for this Figure the full result as calculated in Appendix, to which
  Eq.~(\ref{eq:crossec}) is a good approximation.}
\label{fig:resonances}
\end{figure}

We present a detailed calculation of the squared amplitude
$\frac{1}{4}\sum_{\lambda, spins}|{\cal M}|^2$ in Appendix~\ref{sec:fullcross}. In this section, instead, we focus on those terms that, due to special conditions, feature large enhancements. These terms correspond to: 
\begin{itemize}
\item[(i)] the resonance for $\sqrt{s}=M_{\tilde e}$, i.e.
when the selectron is on-shell; 
\item[(ii)] a logarithmic enhancement, when
the photon is co-linear with the final electron. 
\end{itemize}
The terms in the cross sections that feature both (i) and (ii) are (see Appendix~\ref{sec:fullcross}):
\beqn  \label{eq:leading}
\frac{1}{4}\sum_{\lambda,spins}|\varepsilon_\mu^*(\lambda){\cal M}_2^\mu|^2
&=&2e^2(a_L^4+a_R^4)\frac{|\Pi_s|^2}{t'}(pk)(p'q), \nn\\
\frac{1}{4}\sum_{\lambda,spins}
\left\{\varepsilon_\mu^*(\lambda){\cal M}_2^\mu 
\varepsilon_\nu(\lambda){\cal M}_3^{*\nu}+h.c\right\}
&=&-2e^2(a_L^4+a_R^4)\frac{|\Pi_s|^2}{t'}|\Pi_{s'}|^2
[s'-M_{\tilde e}^2+\frac{\sqrt{s'}\Gamma^2}{\sqrt{s}+\sqrt{s'}}]
(p'k')4(pk)^2.
\eeqn
Note that the enhancement (i) comes from $|\Pi_s|^2$, when
$s=M_{\tilde e}^2$, while (ii) comes from $1/t'$.  Besides (i),
there is another resonance, associated with $\Pi_{s'}$ for
$\sqrt{s'}=M_{\tilde e}$, which depends on the final photon
energy. This latter resonance is always less important than the first
one (see {\it e.g.} Fig. ~\ref{fig:resonances}). 
 Notice that one might expect that we should include the
term $2$Re$[{\cal M}_2{\cal M}_1^*]$, which is also logarithmically
enhanced. However that term features a factor Re($\Pi_s
\Pi_{s'}^*$) which produces a much smaller enhancement than $|\Pi_s|^2$. Thus,
the $2$Re$[{\cal M}_2{\cal M}_1^*]$ term is less important than the ones considered in
Eq.~(\ref{eq:leading}) and we can safely neglect it. Our numerical work
supports this claim, as we will also see {\it e.g.} in Fig.~\ref{fig:compare} below.

We can now compute the overall differential cross section 
\beqn
d\sigma=&&(2\pi)^4\delta^4(p+k-p'-k'-q)\frac{1}{4M_\chi E}
\frac{d^3 k'}{2E'(2\pi)^3}\frac{d^3 q}{2E_\gamma(2\pi)^3}
\frac{d^3 p'}{2E'_N(2\pi)^3}\nn\\
&& \times \frac{1}{4}\sum_{\lambda,spins}\left(|\varepsilon_\mu^*(\lambda){\cal M}_2^\mu|^2+\left\{\varepsilon_\mu^*(\lambda){\cal M}_2^\mu 
\varepsilon_\nu(\lambda){\cal M}_3^{*\nu}+h.c\right\}\right).
\eeqn
We first concern ourselves with calculating the kinematics. We work in the approximation of massless electrons,
except when we calculate logarithmic terms (see Eq.~(\ref{eq:log})).  We assume the neutralino to be at
rest, $p=(M_\chi,0,0,0)$, and we will eventually integrate over the spectrum of
incoming electrons. The quantity that is actually measured, the photon energy, in our notation is $q$. Therefore, we need to integrate over the $p',k'$ phase space. 
The $p'$ integration is done with the $\delta$-function, so only $k'$ remains. 
The scalar products in Eq.~(\ref{eq:leading}),
\beqn
(p'k')&=&(p+k-k'-q,k')\to(pk)-(p+k,q)\nn\\
(p'q)&=&(p+k-k'-q,q)\to(p+k,q)
\eeqn
do not depend on angles involving $k'$. In order to have the logarithmic
enhancement, we want the photon to be co-linear with the final
electron: this corresponds to the so-called co-linear approximation, that will also further
simplify the calculation of the kinematics.
After integrating over $d^3p'$ with the delta function, we are left
with the energy delta function that gives us the on-shell condition for the final electron energy:
$\delta(E_N+E-E_\gamma-E'-E'_N)=\delta(E'+E_\gamma-E_C)$, with the usual 
Compton laboratory-frame energy $E_C=\frac{E}{1+\frac{E}{M_\chi}(1-\cos\theta)}$, $\theta$ 
being the angle between the initial electron and final photon. Finally,
\beqn \label{eq:crossec}
\frac{d^2\sigma}{dE_\gamma
  d\Omega}=&&\frac{1}{(2\pi)^5}\frac{1}{32E'_N}2e^2(a_L^4+a_R^4)|\Pi_s|^2
\nn\\
&& \times
\left[E_\gamma\left(M_\chi+E(1-\cos\theta)\right)-|\Pi_{s'}|^2\left(s'-M_{\tilde
      e}^2+\frac{\sqrt{s'}\Gamma^2}{\sqrt{s}+\sqrt{s'}}\right)4EM_\chi\left(EM_\chi-E_\gamma\left(M_\chi+E(1-\cos\theta)\right)\right)\right]\nn
\\
&& \times \int d\Omega_{k'} \frac{E_\gamma E'}{t'}
\eeqn
and the integral over the solid angle can easily be evaluated (here one has to keep the 
electron mass finite) to give
\beqn \label{eq:log}
\int d\Omega_{k'}\frac{E_\gamma E'}{t'}=\pi\ln\left(\frac{4E'^2}{m_e^2}\right).
\eeqn
Above, $E'_N=M_\chi+\frac{E^2(1-\cos\theta)}{M_\chi+E(1-\cos\theta)}$
is the energy of the neutralino in the final state.
Note that from the delta function we have the condition 
$E'+E_\gamma=\frac{E}{1+\frac{E}{M_\chi}(1-\cos\theta)}$ which leads to a 
lower limit on the initial electron energy for a given angle between the jet and the observer, $\cos\theta$, and outgoing photon energy
$E_\gamma$:
\beqn \label{eq:Emin}
E_{min}=\frac{E_\gamma}{1-\frac{E_\gamma}{M_\chi}(1-\cos\theta)}.
\eeqn

A few comments are in due order at this point:
\begin{itemize}
\item For a given photon kinematics, only those particles in the initial electron 
spectrum with $E\geq E_{min}$ contribute.
\item From Eq.~(\ref{eq:Emin}) we see that the photon energy cannot exceed $\frac{M_\chi}{1-\cos\theta}$.
\item The cross section in Eq.~(\ref{eq:crossec}) contains two
  resonances: one for $s=M_{\tilde e}^2$, that doesn't depend on
  $E_\gamma$, and a second one for $s'=M_{\tilde e}^2$, that instead does depend on
  $E_\gamma$. The latter, however, gives a peak which is always significantly
  lower than the former.
\item When we scan over photon energies, $E_{min}$ gets larger as
  $E_\gamma$ increases. Eventually $E_{min}$ gets big enough that
  $s>M_{\tilde e}^2$ and we lose the first resonance. This happens
  when
\be \label{eq:drop}
E_\gamma = \frac{M_\chi (M_{\tilde e}^2 -
  M_\chi^2)}{2M_\chi^2+(M_{\tilde e}^2 - M_\chi^2)(1-\cos\theta)}.
\ee
\end{itemize}
Some of these features are illustrated in the plots of Fig.~\ref{fig:resonances}, where we show the full differential cross section of
  Eq.~(\ref{eq:fullcross}) as a function of the initial electron energy
  $E$ for different values of the final photon energy $E_\gamma$ (not the approximated expression we give in Eq.~(\ref{eq:crossec})).
  In the first plot, where we set $E_\gamma=10$ GeV we see the two resonances corresponding to $s=M_{\tilde
  e}^2$ and $s'=M_{\tilde e}^2$. In the second plot, where $E_\gamma=30$ GeV, we still have both
resonances, but the line starts at a larger value of $E=E_{min}$
because we of the increased value of $E_\gamma$. Finally, in the third plot, with $E_\gamma=50$ GeV, $E_{min}$ is
above the value needed for the first resonance to occur, and we only see the
second one. As explained above, the first resonance
does not move when we vary $E_\gamma$, while the second one does. Note
also that, when present, the first peak is much higher than the second
one. For all three plots we fixed the following parameters: $M_{\tilde e}=100$
GeV, $M_\chi = 60$ GeV, $\theta=68^\circ$ (the latter angle corresponding to the line-of-sight angle of the Cen A jet).

The Reader may notice the apparent difference between the present calculation and the one presented in the work of Bloom and Wells, Ref.~\cite{Bloom:1997vm}: while the result of Ref.~\cite{Bloom:1997vm} is of order $a_{L,R}^2$, ours is $\sim a_{L,R}^4$. However, 
this is only seemingly a discrepancy because the two extra powers of 
$a_{L,R}$ are cancelled by the selectron width in the denominator ($\tilde 
e\to e+\chi$ is the only decay channel) -- see e.g. Eq.~(\ref{eq:deltaf}) below. This further leads to the 
resonant enhancement $\sim1/(m_{\tilde e}^2-m_\chi^2)$. This mechanism, 
along with the log enhancement $\sim\ln(E^2/m_e^2)$, is absent in Bloom and Wells \cite{Bloom:1997vm}.

\subsection{Lightest KK particle Dark Matter}
\begin{figure}[!t]
\centering
\subfigure[${\cal M}_1$]{\label{fig:UED1} \includegraphics[width=50
  mm]{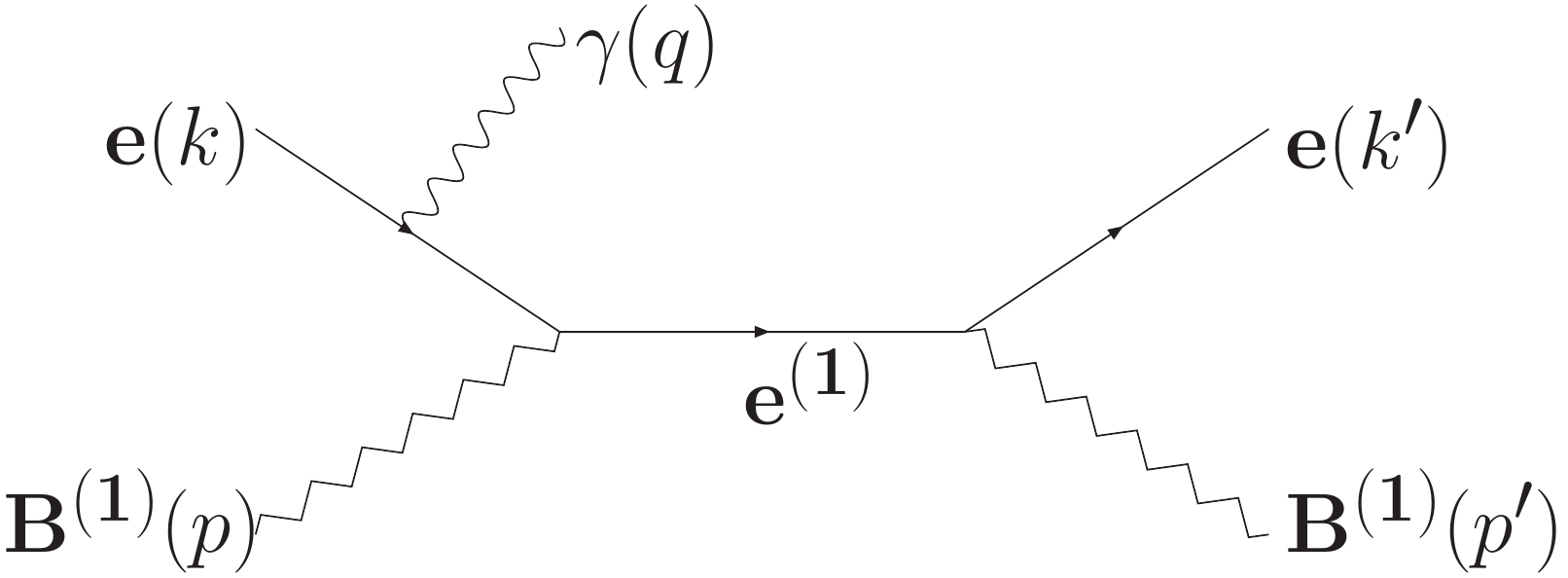} }
\hspace{5 mm}
\subfigure[${\cal M}_2$]{\label{fig:UED2} \includegraphics[width=50
  mm]{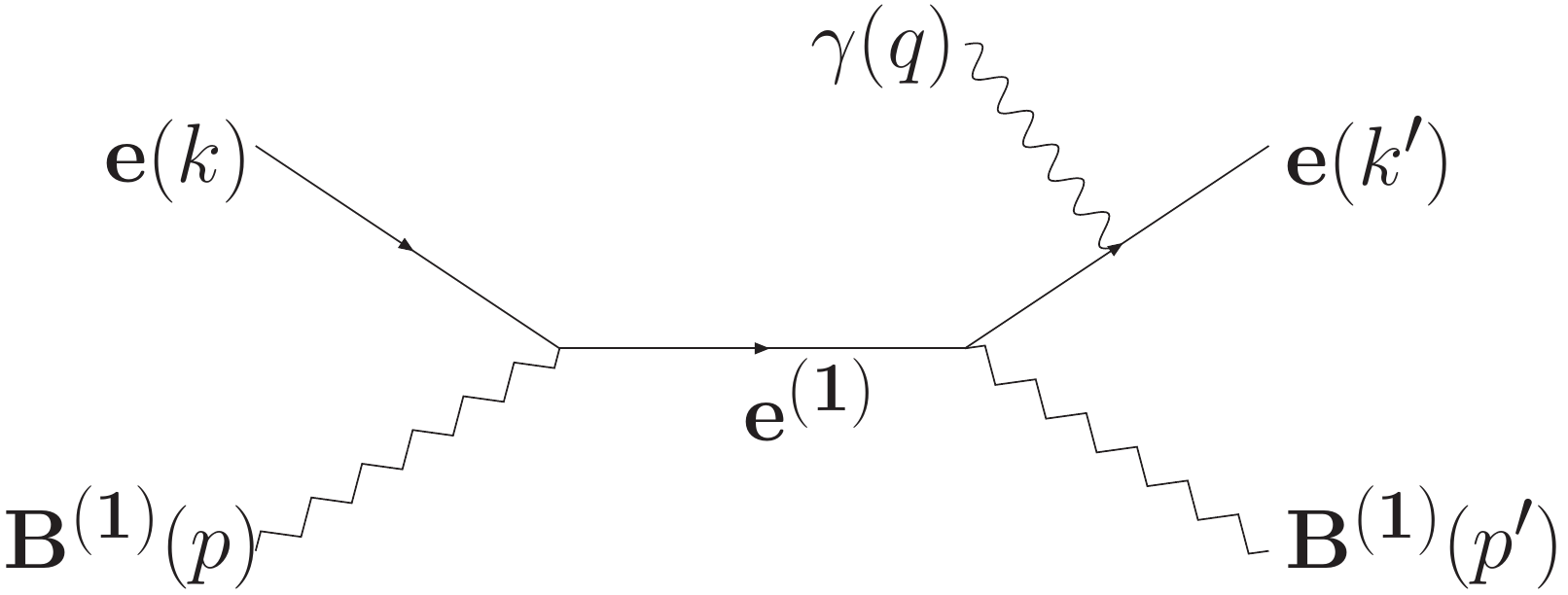}  }
\hspace{5 mm}
\subfigure[${\cal M}_3$]{\label{fig:UED3} \includegraphics[width=50
  mm]{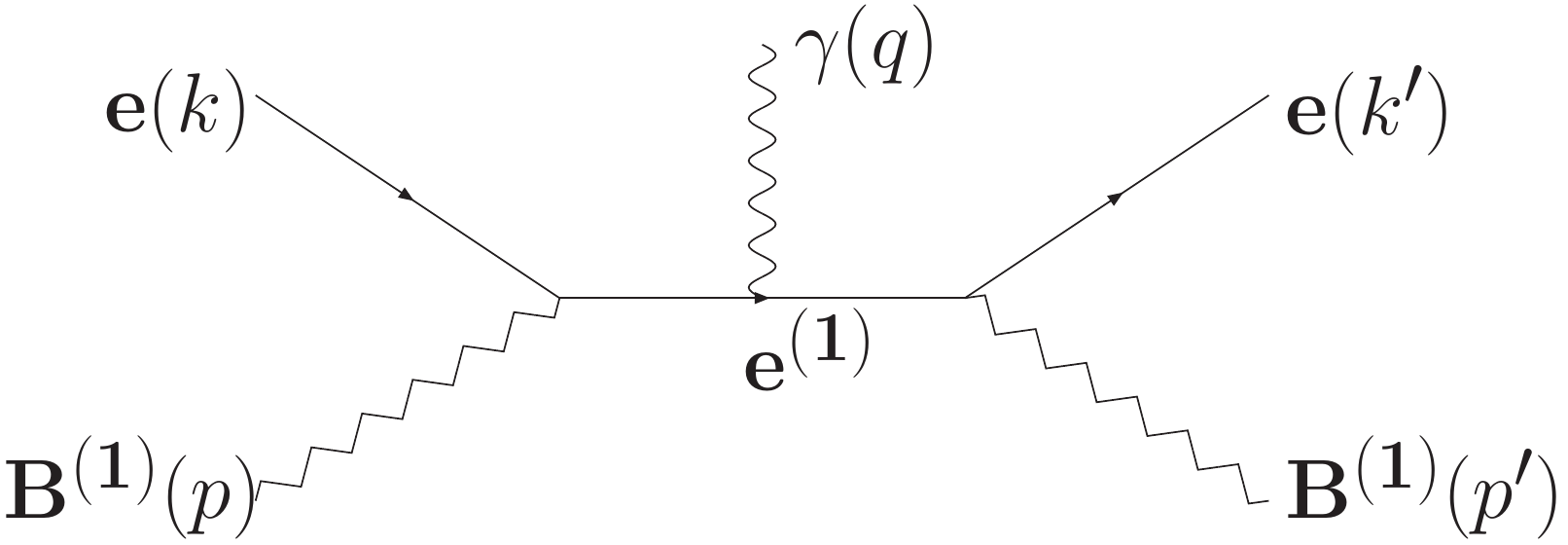} }
\caption{Feynman diagrams for the process $e(k)+B^{(1)}(p) \rightarrow e(k')+B^{(1)}(p')+\gamma(q)$.}
\label{fig:UEDdiagrams}
\end{figure}

With the insight gained from the calculation of the full differential cross section for the supersymmetric case, we can now turn to the case of UED, where the dark matter candidate is a massive vector boson corresponding to the Kaluza-Klein first excitation of the hypercharge gauge boson, $B^{(1)}$, a particle that is also assumed to be the Lightest KK Particle (LKP) \cite{uedrev}. The relevant Feynman diagrams for this model, where the scalar-electron is replaced by the Kaluza-Klein electron, are shown in Fig.~\ref{fig:UEDdiagrams}. 

We define
\beqn
M&\equiv&M_{B^{(1)}}\nn\\
\delta&\equiv&M_{e^{(1)}}-M_{B^{(1)}}<<M.
\eeqn
Note that we assume the KK first modes to be nearly degenerate \cite{uedrev}.
Under this assumption, we have a very small width\footnote{$Y_L$($Y_R$) is the hypercharge of the left-(right-)handed electron, $g_1$ is the hypercharge gauge coupling.} $\Gamma=\frac{3}{8\pi}\frac{\delta^2}{M}g_1^2(Y_L^2+Y_R^2)$ for the decay of the $e^{(1)}$, so we are looking for a $\delta$-like peak in our
calculation, and we can perform the following substitution:
\beqn\label{eq:deltaf}
|\frac{2M_{e^{(1)}}}{P^2-M_{e^{(1)}}^2-iM_{e^{(1)}}\Gamma}|^2\;=\;\frac{1}{(E-\delta)^2+\Gamma^2/4}\to\frac{2}{\pi\Gamma}\delta(E-\delta).
\eeqn
To compute the differential cross section, we can apply here the same line of reasoning as for the supersymmetric case. Referring to the diagrams of Fig.~\ref{fig:UEDdiagrams}, we only need to compute $\frac{1}{6}\sum_{\lambda,spins}|{\cal M}_2|^2$ and
$\frac{1}{6}\sum_{\lambda,spins}
\left\{{\cal M}_2{\cal M}_3^*+h.c\right\}$ for the leading contribution. We have
\beqn
{\cal M}_2^{\mu\nu\alpha}&=&ieg_1^2\bar
u(k')\gamma^\alpha\frac{\sk'+\sq}{(k'+q)^2-m_e^2}
\gamma^\nu\frac{Y^2_\Lambda(\sp+\sk)+Y_LY_RM_{e^(1)}}{(p+k)^2-M_{e^(1)}^2-iM\Gamma}\gamma^\mu
P_\Lambda u(k)\nn\\
&=&\frac{ieg_1^2}{2t'(E-\delta-i\Gamma/2)},u(k')(2k'^\alpha+\gamma^\alpha\sq)
\gamma^\nu[Y^2_\Lambda\sv+Y_LY_R]\gamma^\mu P_\Lambda u(k),\nn\\
{\cal M}_3^{\mu\nu\alpha}&=&ieg_1^2\bar u(k') (Y_RP_L+Y_LP_R)
\gamma^\nu \frac{\sp'+\sk'+M_{e^(1)}}{(p'+k')^2-M_{e^(1)}^2-iM\Gamma}
\gamma^\alpha \frac{\sp+\sk+M_{e^(1)}}{(p+k)^2-M_{e^(1)}^2-iM\Gamma}
\gamma^\mu
(Y_RP_R+Y_LP_L) u(k)\nn\\
&=&
\frac{ieg_1^2}{4(E-E_\gamma-\delta-i\Gamma/2)(E-\delta-i\Gamma/2)}\bar u(k')(Y_RP_L+Y_LP_R)
\gamma^\nu (\sv +1)\gamma^\alpha(\sv+1)
\gamma^\mu (Y_RP_R+Y_LP_L) u(k), \nn \\
\eeqn
with $\Lambda=R,L$ the chirality of the electron, and
$v^\mu=\frac{p^\mu}{M}=(1,\vec 0)$ in the ``laboratory'' frame. The resulting spin-averaged squared amplitudes are:
\beqn
\frac{1}{6}\sum_{\lambda, spins}|{\cal
  M}_2|^2&=&\frac{3}{2}e^2g_1^2(Y_L^2+Y_R^2)^2\frac{E E_\gamma}{t'[(E-\delta)^2+\Gamma^2/4]}
\left[1+\frac{1}{9}\frac{(Y_L^2-Y_R^2)^2}{(Y_L^2+Y_R^2)^2}\cos\theta\right], \nn \\
\frac{1}{6}\sum_{\lambda, spins} 2{\rm Re}[{\cal M}_2{\cal M}^*_3]&=&-3e^2g_1^4(Y^2_L+Y^2_R)^2\frac{(E-E_\gamma-\delta)EE'(E'+E_\gamma)}{t'[(E-\delta)^2+\Gamma^2/4][(E-E_\gamma-\delta)^2+\Gamma^2/4]}\left[1+\frac{1}{9}\frac{(Y_L^2-Y_R^2)^2}{(Y_L^2+Y_R^2)^2}\cos\theta \right]\nn\\
\eeqn
Note that we have two scales in this calculation, $E\sim E' \sim E_\gamma \sim \delta \ll M$. To derive the Equations above we dropped terms proportional to $\delta/M$. The kinematics is the same as in the neutralino case. Because of the hierarchy between the two scales, the calculation is here even  simpler. For instance, the Compton laboratory energy is $E_C\simeq E$, the energy of the final $B^{(1)}$ is $E'_B \simeq M$ and the lower limit onto the initial electron energy just reduces to $E_{min} \simeq E_\gamma$. Finally, the differential cross section for the UED $B^{(1)}$ case reads:
\be
\frac{d^2\sigma}{dE_\gamma d\Omega_q}=\frac{e^2 g_1^4}{(2\pi)^5}\frac{3\pi}{64 ME'_B}\frac{(Y_L^2+Y_R^2)^2+\frac{1}{9}(Y_L^2-Y_R^2)^2\cos\theta}{(E-\delta)^2+\Gamma^2/4}\left[E_\gamma - 2\frac{(E-E_\gamma-\delta)E'(E'+E_\gamma)}{(E-E_\gamma-\delta)^2+\Gamma^2/4} \right] \ln\left(\frac{4E'}{m_e^2}\right).
\ee

\section{Photon flux} \label{sec:flux}

\begin{figure}
\centering
\includegraphics[width=130mm]{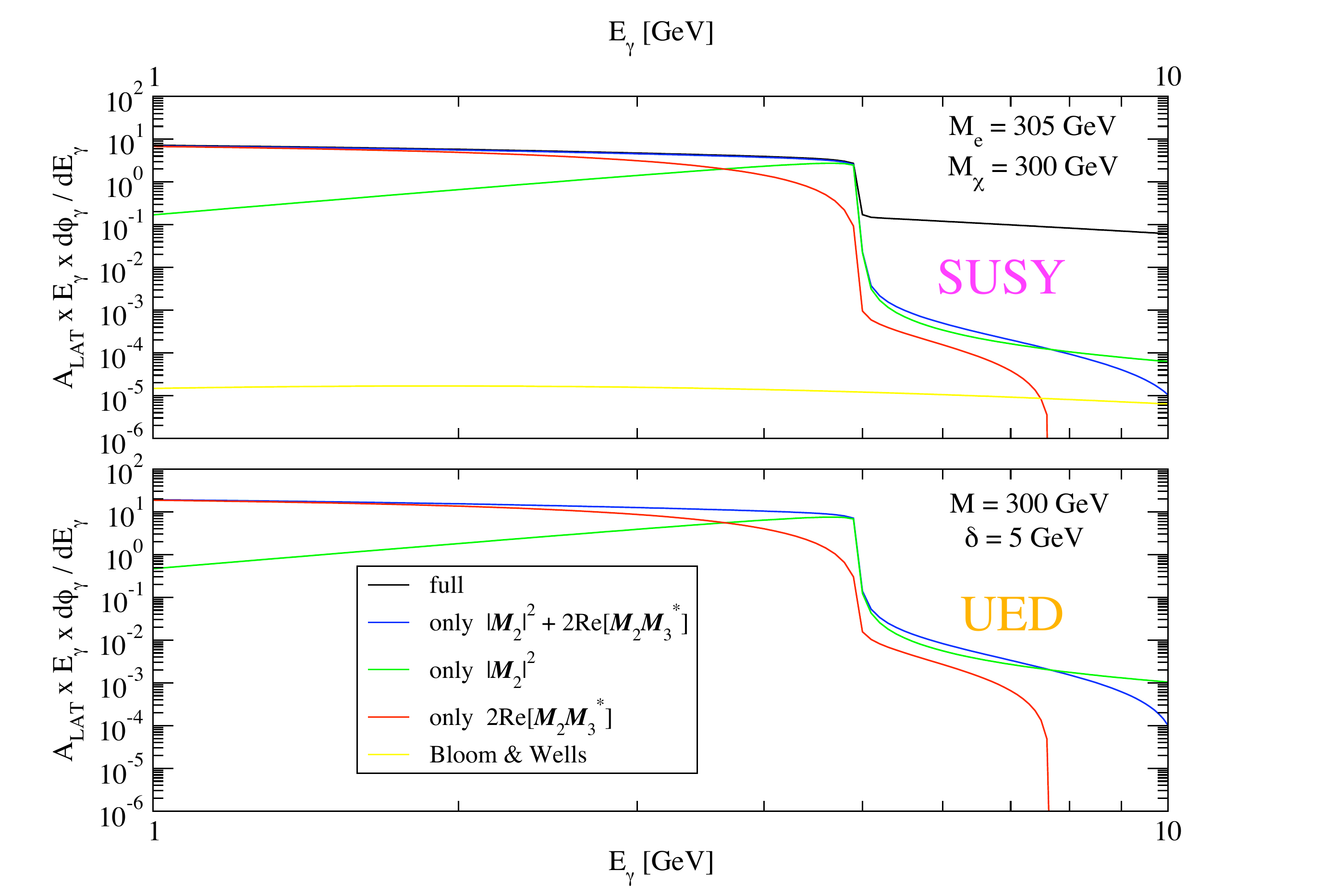}
\caption{We plot $A_{\rm LAT}E_\gamma \frac{d\phi_\gamma}{dE_\gamma}$
  vs $E_\gamma$ for the two scenarios we studied: neutralino dark
  matter (top) and LKP dark matter (bottom). To obtain the black line
  in the top plot we used the full result Eq.~(\ref{eq:fullcross}) for
  the differential cross section. Note how the blue line coincides
  with the black one until the curves drop down. This confirms the
  validity of the approximation we made in the calculation of the
  differential cross section, namely that the main contribution comes
  from $|{\cal M}_2|^2$ and 2Re$[{\cal M}_2{\cal M}_3^*]$. As
  expected, the curves drop down when we lose the main resonance in
  the cross section (see Eq.~(\ref{eq:drop}) for the SUSY case). The
  two cases give an almost identical signal, apart from a constant
  that can be traced back to the different coupling constants. In the
  top plot we also show with a yellow line the signal one would obtain considering only
  the process with no resonances studied originally in Ref.~\cite{Bloom:1997vm}. The
  angle $\theta$ is here fixed at 68$^\circ$, a value appropriate
  for Cen A.}
\label{fig:compare}
\end{figure}

We now have all the ingredients to compute the differential flux of Eq.~(\ref{eq:flux}), with educated guesses for the two AGNs Centaurus A and M87, and we can thus proceed to compare the process under investigation here with actual gamma-ray observations. Prior to comparing with data, we carry out a parallel study of the supersymmetric versus UED dark matter paradigms as far as e[p]-dark matter scattering is concerned. In Fig.~\ref{fig:compare} we  plot the results for the supersymmetric case (top) and the UED case (bottom), for comparable values of the mass parameters (dark matter particle mass set to around 300 GeV, and mass splitting between the dark matter particle and the intermediate state scalar-electron or Kaluza-Klein electron of 5 GeV; notice that in the case of supersymmetry while such a small mass splitting might not be natural, it is sometimes warranted by the requirement of a co-annihilation amplitude to enforce the correct thermal relic abundance for the lightest neutralino). The yellow curve shows the full result for the $e\chi\to \tilde e \gamma$ amplitudes considered in Ref.~\cite{Bloom:1997vm} (not the approximated estimate to the cross section given there) i.e. with the final selectron on shell, and illustrates one of the reasons why the final fluxes calculated here are significantly larger than those presented in that study. 

Note that the curves for both UED and supersymmetry are very similar. The fact that the UED curve is slightly larger is accidental, and due to the different values of the couplings at the vertices of the Feynman diagrams and not to the intrinsic structure of the relevant scattering process. From the plot at the top we also have an explicit confirmation of the line of reasoning  we outlined in the previous section: the dominant contribution to the process of interest here stems from $|{\cal M}_2|^2$ and 2Re$[{\cal M}_2{\cal M}_3^*]$. The other contributions are not important until photon energies become greater than 5 GeV. As we can see, however, the signal drops anyway in that region, so those contributions are never crucial.

\begin{figure}
\centering
\includegraphics[width=170mm]{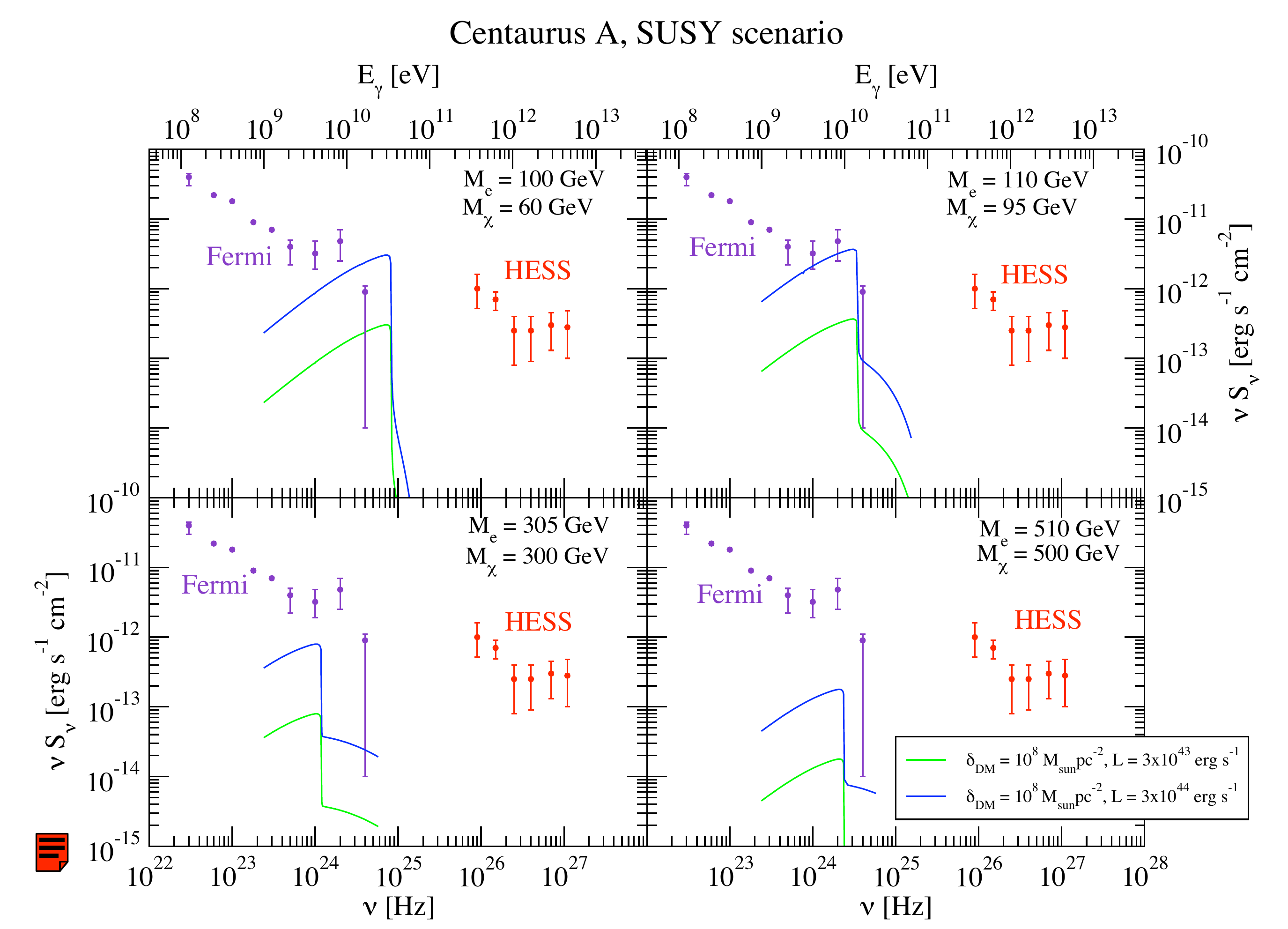}
\caption{We show the spectral energy distribution $\nu S_\nu$
  (equivalent to $E_\gamma^2 \times \frac{d\Phi_\gamma}{dE_\gamma}$)
  versus the photon energy for CenA, for four different choices of the neutralino and lightest selectron
  mass in a supersymmetric scenario.  As explained in the text
  astrophysical uncertainties affect both the value of the
  integral $\delta_{\rm DM}$ and the jet luminosity. Fermi-LAT data are from Ref.~\cite{fermicena}. The H.E.S.S. data are from Ref.~\cite{Aharonian:2009xn}.}
\label{fig:CSUSYconvolution}
\end{figure}

We are now in the position to answer the question of whether current
experiments might detect a signal coming from the process under present investigation. We first consider the case of Cen A \cite{fermicena}, and we start focusing on the supersymmetric case. In Fig.~\ref{fig:CSUSYconvolution} we plot the spectral energy distribution, i.e. $E_\gamma^2 \times
\frac{d\Phi_\gamma}{dE_\gamma}$ versus the photon energy, and we compare with 
Fermi-LAT \cite{fermicena} and H.E.S.S. \cite{Aharonian:2009xn} observations. As explained in previous sections, it is reasonable
to consider a fairly generous range of plausible values for the $\delta_{\rm DM}$
integral as well as for the luminosity of the jet, due to
uncertainties in the physics of the AGN jet and on the dark matter density distribution in the innermost AGN regions. In
Fig.~\ref{fig:CSUSYconvolution} we show lines corresponding to two different
choices for the jet luminosity, that given the results of \cite{fermicena} can be regarded as relatively conservative, and we use the most conservative value for $\delta_{\rm DM}$, corresponding to the black line in Fig.~\ref{fig:Cprofiles}.

The particle dark matter mass is set to 60, 95, 300 and 500 GeV, from top-left to bottom-right, with a range of mass splittings between the selectrons and neutralino. Intriguingly enough, we find that the resulting gamma-ray flux is detectable for small enough values of the mass (less than about 300 GeV) and a luminous enough jet, even for the most conservative assumption on the dark matter density distribution. The spectral structure of the predicted flux of photons from electron-dark matter scattering is very hard and has a dramatic drop-off at the edge of the resonance. The location of such a sharp break is between 5 and 50 GeV for the models under consideration, the ideal energy range for Fermi-LAT observations.  We also find that no photons are predicted to be observable by Atmospheric Cherenkov Telescopes (ACTs) such as H.E.S.S. It is somewhat fascinating that the Fermi-LAT data from the Cen A system do show a hardening of the spectrum right before an energy of 10 GeV.

Models where the mass splitting between the selectron and the neutralino is much larger than the ones considered, are not very interesting here. In those cases, the drop-off of the photon flux happens at higher energy, and one might hope to get in the region probed by H.E.S.S. The signal would be too low to be detectable though, mostly due to the fact that, in order to hit the resonance, we would now need electrons with much higher energies, that would put us in the tail of the electron energy distribution.

Given the very large values of dark matter density, a potential additional source of gamma rays is direct dark matter annihilation close to the AGN center. The flux of such photons can be estimated as
\be
\left(\frac{d\Phi}{dE_\gamma} \right)_{\rm annihilation} =\frac{{\rm d}N_\gamma}{{\rm d}E_\gamma} \frac{\langle \sigma v \rangle_0}{2M_\chi^2 4\pi d^2_{\rm AGN}}\int_{r_{min}}^{r_0} dr \ 4\pi r^2 \rho^2_{\rm DM}(r),
\ee
where the factor ${\rm d}N_\gamma/{\rm d}E_\gamma$ indicates the differential photon spectrum per annihilation event for the particular dark matter model under consideration. Notice that the flux above (and in particular the spectrum encoded in ${\rm d}N_\gamma/{\rm d}E_\gamma$) is entirely disconnected (in a model independent way) from the flux originating from e[p]-dark matter scattering we discuss here. In principle, one can have the latter without having any flux from the former. Indeed, if we consider a small enough annihilation cross section, $\langle \sigma v
\rangle_0 \sim 10^{-30}$ cm$^3$s$^{-1}$, typical values of $E^2_\gamma \left(\frac{d\Phi}{dE_\gamma} \right)_{\rm annihilation}$
will be  $\sim 10^{-13}$, $10^{-11}$ erg s$^{-1}$ cm$^{-2}$. These are
comparable to the photon fluxes shown in
Fig.~\ref{fig:CSUSYconvolution}, coming from the scattering of
electrons off the neutralinos. Notice that the signal shown in the Figure is
peaked at $E_\gamma \sim$ a few GeV, for most of the cases considered,
whereas the signal from dark matter annihilation would be peaked at
$E_\gamma \lesssim M_\chi$, so, even if the values of the fluxes were
similar, we would see two distinct signatures. Also, in scenarios where
the neutralino and the selectron are nearly degenerate, the
coannihilations of dark matter with selectrons will favor a small value for the
pair annihilation cross section. This, in turn, would enhance the
value of the $\delta_{\rm DM}$ and, therefore, the flux of photons
from the scattering process we studied, while it would result in a less
important flux for photons coming from the pair annihilation.

\begin{figure}
\centering
\includegraphics[width=170mm]{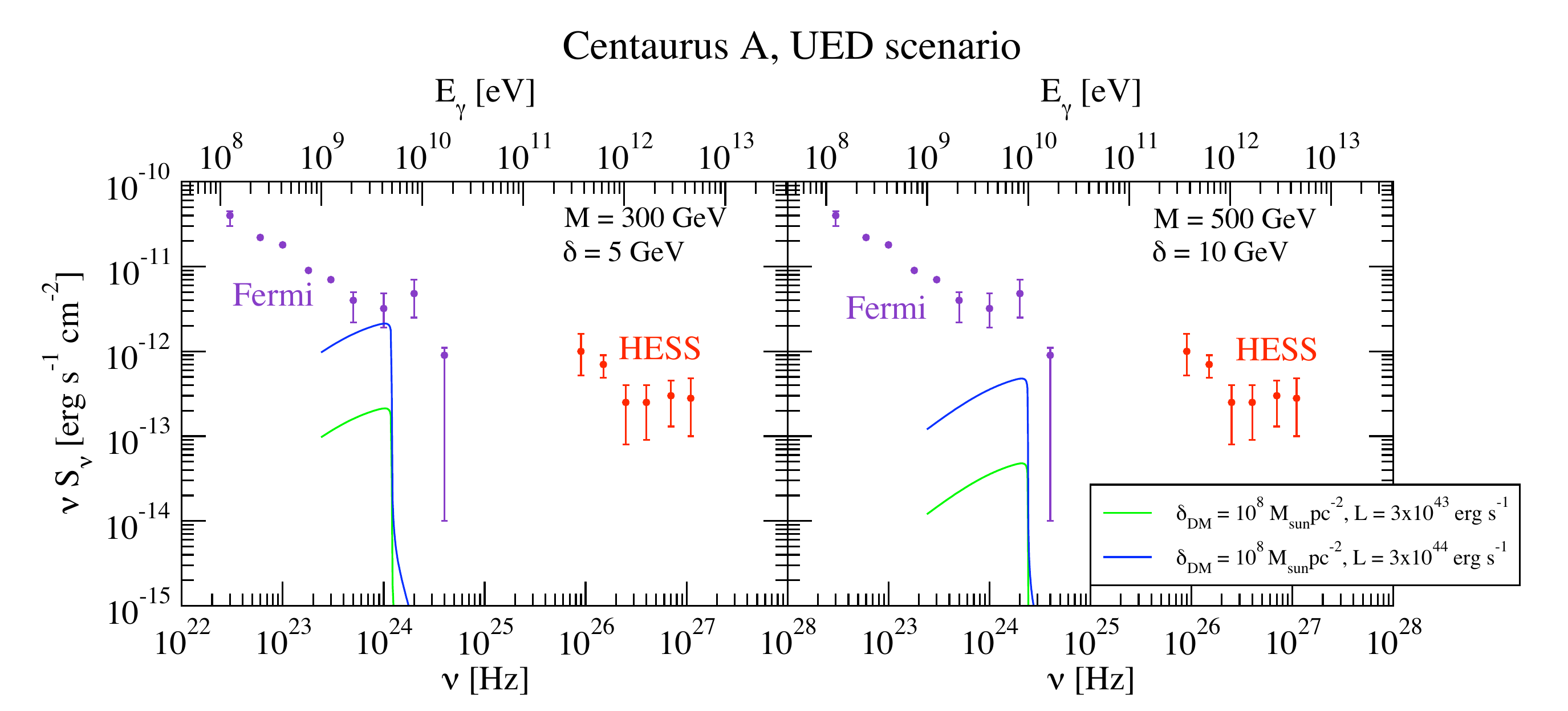}
\caption{We show the spectral energy distribution $\nu S_\nu$
  (equivalent to $E_\gamma^2 \times \frac{d\Phi_\gamma}{dE_\gamma}$)
  versus the photon energy for CenA, for two different choices of the
  masses in a UED scenario. Fermi data are from Ref.~\cite{fermicena}, while H.E.S.S. data are from Ref.~\cite{Aharonian:2009xn}.}
\label{fig:CUEDconvolution}
\end{figure}

In Fig.~\ref{fig:CUEDconvolution} we show the results for the UED case, again for the Cen A system, and for the same choices of AGN parameters as for the previous Figure. This time the dark matter mass was set to 300 (left) and 500 GeV (right), with a mass splitting of 5 (left) and 10 GeV (right), close to what expected in the context of UED \cite{uedrev}. Thanks to a larger scattering cross section, the UED case gives a potentially detectable signal, for a large enough jet luminosity, even with a dark matter mass of 300 GeV.  

The next Figure~\ref{fig:cenabestfit} shows how the Fermi-LAT data from Cen A in the few GeV range can be in principle fitted with the emission originating from electron-dark matter scattering, for selected values of the particle dark matter mass, and of the astrophysical parameters $\delta_{\rm DM}$ and the jet luminosity $L$, for both supersymmetry (left) and UED (right). In both panels, we set the Cen A jet luminosity to the somewhat large value of $3\times10^{44}$ erg/s. This value can be partly traded off for a larger value of $\delta_{\rm DM}$, which we fix here to lie towards the lower end of the expected range (see Fig.~\ref{fig:Cprofiles}) to $\delta_{\rm DM}=2\times 10^8$ (left) and $6\times 10^8$ (right) $M_\odot/{\rm pc}^2$. The predicted spectral drop-off due to the location of the resonance (set by the splitting between the dark matter particle and the intermediate charged particle) follows the Fermi data for the two particular choices we made for the dark sector spectrum: 95 GeV for the neutralino mass and 110 GeV for the selectron in the case of supersymmetry, and 300 and 315 GeV for the Kaluza-Klein ``photon'' and electron for UED.
\begin{figure}
\centering
\includegraphics[width=170mm]{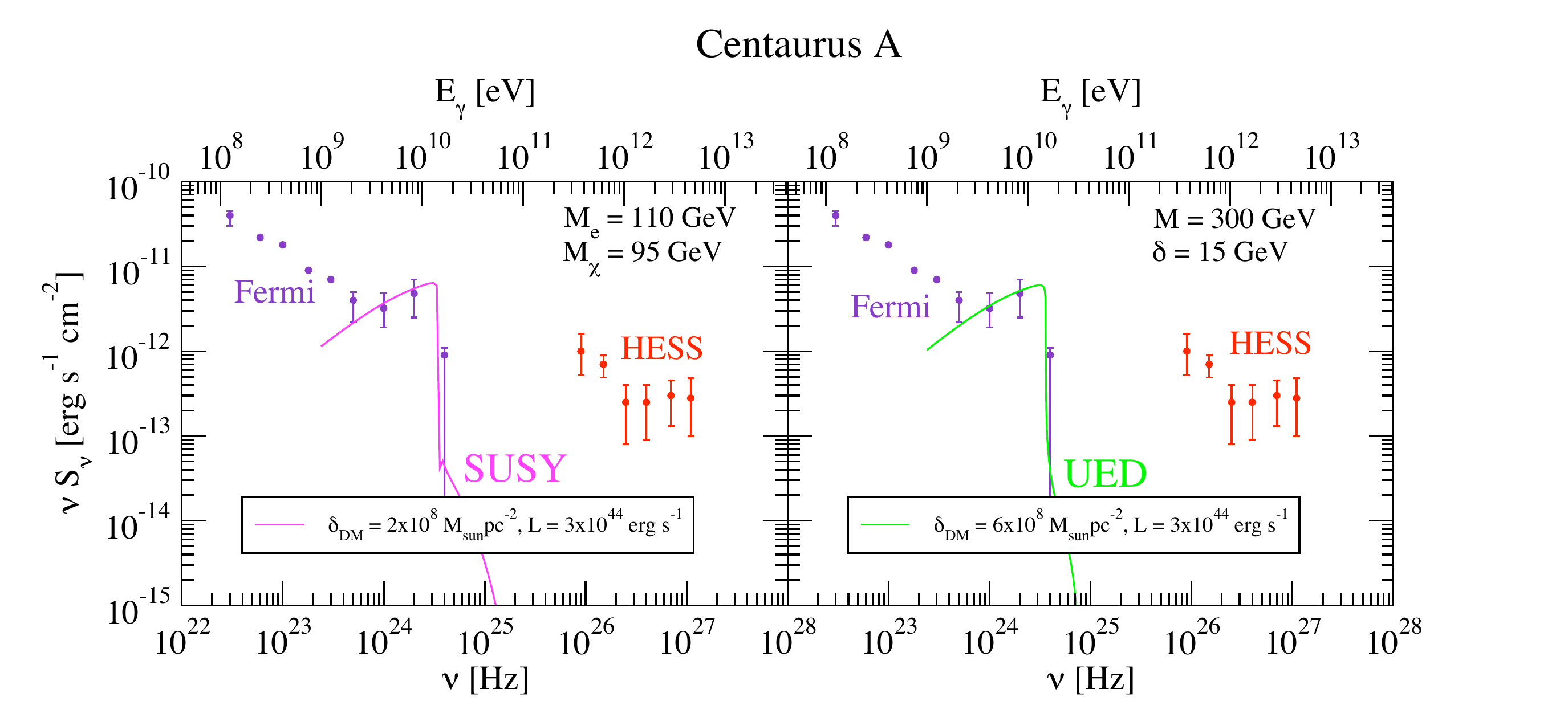}
\caption{We illustrate in this Figure that, with astrophysical parameters adjusted ad hoc, one
  can fit the data from Fermi-LAT for appropriate masses within both a supersymmetric (left) and universal-extra-dimensional (right) dark matter model.}
\label{fig:cenabestfit}
\end{figure}

\begin{figure}
\centering
\includegraphics[width=170mm]{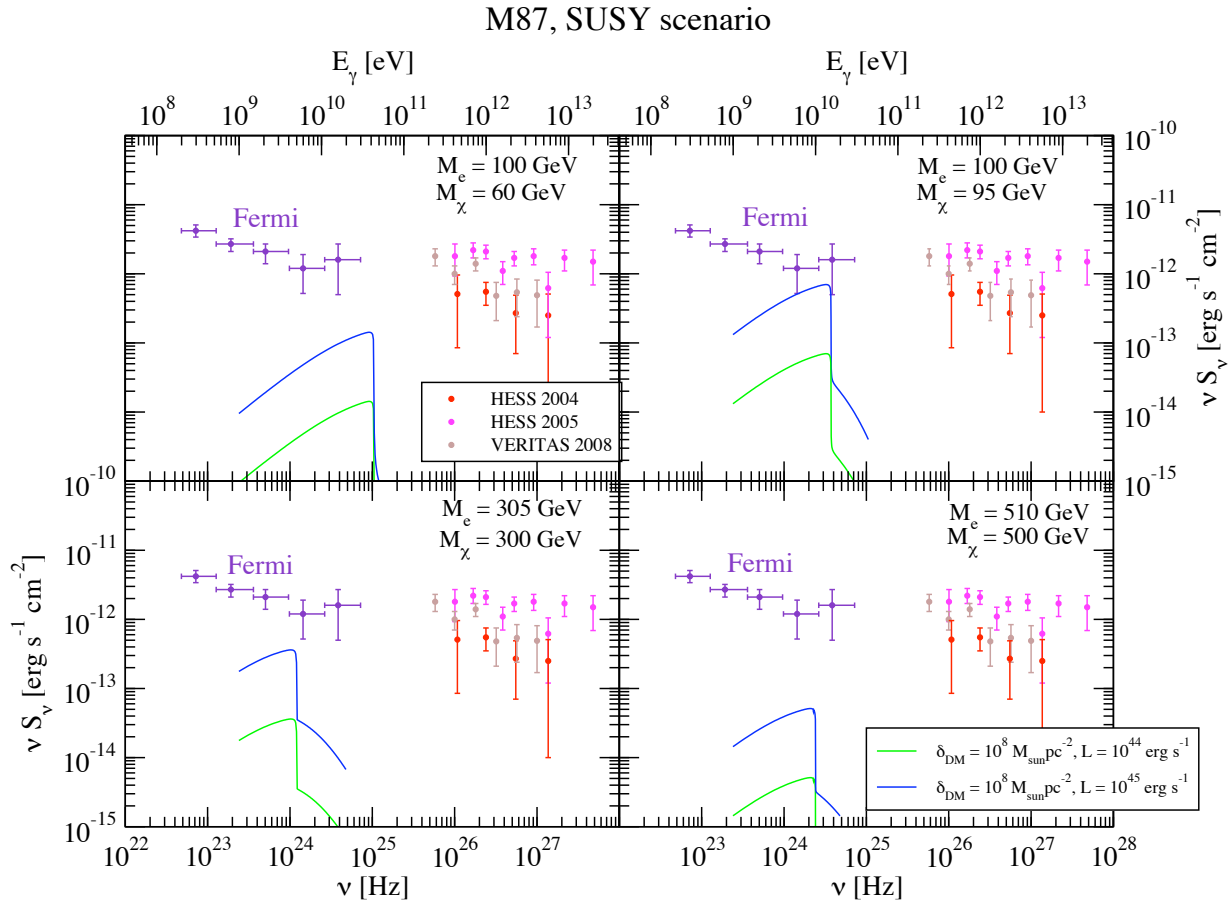}
\caption{ The spectral energy distribution versus the photon energy for M87, for four different choices of the
  masses in a SUSY scenario. The Fermi data are from Ref.~\cite{fermim87}. The H.E.S.S. 
measurements\cite{Aharonian:2006ys} are relative to a low state in
2004 (red) and to a high state in 2005 (magenta). The VERITAS measurements are from Ref.~\cite{Acciari:2010uh}.}
\label{fig:MSUSYconvolution}
\end{figure}

We move on to the case of the M87 jet in Fig.~\ref{fig:MSUSYconvolution}, where we show the results for the SUSY case, for identical values of the neutralino and selectron masses as in the analogue Cen A case shown in Fig.~\ref{fig:CSUSYconvolution}. We compare the predictions for the electron-dark matter scattering discussed here with data from Fermi-LAT \cite{fermim87} as well as from H.E.S.S \cite{Aharonian:2006ys} and VERITAS \cite{Acciari:2010uh} (we do not show here comparable data from the Magic Telescope, see \cite{Albert:2008kb}). The H.E.S.S. measurements we show are relative to a low state in
2004 (red) and to a high state in 2005 (magenta). In the case of M87, we fix $\delta_{\rm DM}$ to $10^8$ $M_\odot$ pc$^{-2}$, and we show two values for the jet luminosity, respectively $10^{44}$ and $10^{45}$ erg s$^{-1}$. Only for a very favorable dark matter spectrum, with the dark matter particle featuring a mass of 95 GeV and the selectron mass at 100 GeV, do we find that the process we consider here (again, for the assumed values of the dark matter density profile and jet luminosity) is relevant with respect to the observed gamma-ray intensity in the Fermi energy range. Also in analogy with the case of Cen A, the spectral drop-off occurs at energies much smaller than those probed by ACT's.

\begin{figure}
\centering
\includegraphics[width=170mm]{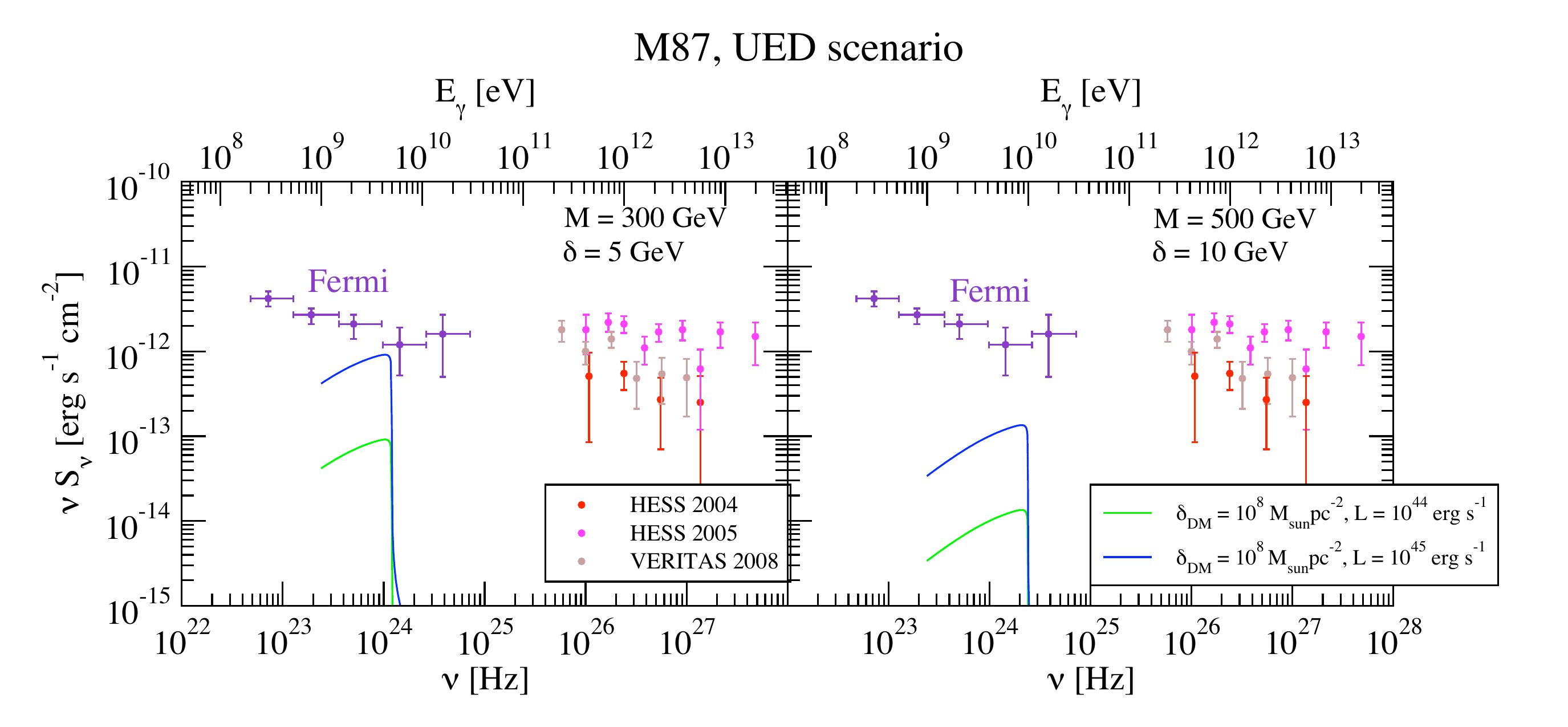}
\caption{ We show the spectral energy distribution $\nu S_\nu$
  (equivalent to $E_\gamma^2 \times \frac{d\Phi_\gamma}{dE_\gamma}$)
  versus the photon energy for M87, for two different choices of the
  masses in a UED scenario. As explained in the text, due to
  astrophysical uncertainties, one can play with the value of the
  integral $\delta_{\rm DM}$ and the jet luminosity within reasonable
ranges. The Fermi data are from Ref.~\cite{fermim87}. The H.E.S.S.
measurements\cite{Aharonian:2006ys} are relative to a low state in
2004 (red) and to a high state in 2005 (magenta). The VERITAS measurements are from Ref.~\cite{Acciari:2010uh}.}
\label{fig:MUEDconvolution}
\end{figure}

Finally, in Fig.~\ref{fig:MUEDconvolution} we show the results for the UED case, for M87, with the same jet and dark matter density profile parameters as for the previous Figure, and with the same particle dark matter setup as in Fig.~\ref{fig:CUEDconvolution}. The electron-LKP scattering process does give rise here to a detectable signal, only slightly smaller than the actual Fermi measurement at an energy of 10 GeV, for an LKP mass of 300 GeV and a mass splitting with the KK electron of 5 GeV. Increasing the mass to 500 GeV reduces the signal significantly (right panel).

In conclusion, we find that for reasonable assumptions on the AGN jet parameters and on the dark matter density distribution around the central massive object of the Centaurus A and M87 AGNs, the Fermi Telescope is in principle sensitive to the detection of photons resulting from the scattering of AGN jet electrons off of dark matter particles. This conclusion holds for specific choices of the dark sector particle spectrum, but is general enough to include both the case of supersymmetric and universal extra dimensional dark matter.


\subsection{Gamma Ray absorption}
Electron-positron pair production in the process $\gamma\gamma\to e^+ e^-$ limits the minimum radius for a compact source below which the source is effectively obscured: gamma rays are lost to electron-positron pair production as they traverse photon fields surrounding the compact source. Quantitatively, this occurs if $\tau_{\gamma\gamma}\approx n_\gamma\sigma_{\gamma\gamma}R\gtrsim1$, where $n_\gamma$ is the target photon number density, $\sigma_{\gamma\gamma}$ is the relevant pair-production cross section, and $R$ is the distance from the source. Indicating with $L_\gamma$ the source luminosity at the energy $E$ corresponding to the peak of the $\gamma\gamma\to e^+ e^-$ cross section for a given gamma-ray energy, we have that this condition translates into the following requirement on the photon density:
\begin{equation}
n_\gamma\approx\frac{L_\gamma}{4\pi R^2 E c},
\end{equation}
where the numerical coefficient in front of the Equation above depends
on the emission geometry. For the gamma ray energies of interest to
us, $E$ lies in the range between 0.1 and 10
keV. Ref.~\cite{1995ApJ...449..105K} estimates that the total
luminosity from the nuclear region of Cen A is $\approx 10^{43}$
ergs/s. This implies that the source is transparent to gamma rays in
the energy range of interest to us, since the radius $R$ at which
$\tau_{\gamma\gamma}\sim1$ is much smaller than $R_S$. In the case of
M87, instead, Ref.~\cite{1998ApJ...493L..83T} finds that integrating the
continuum spectrum of the nuclear region of M87 between 100 $\mu$m ($\sim1.2\times 10^{-5}$ keV) and
10 keV the total luminosity in the highest state is on the order of
$3\times 10^{42}$ ergs/s. Since $R_S$ is much larger for M87 than for Cen A,
this implies that also for M87 the in-situ absorption of gamma rays is
negligible for the radii of interest for the present study.

\section{Proton Jets}\label{sec:protons}
\begin{figure}[!t]
\centering
\includegraphics[width=50mm]{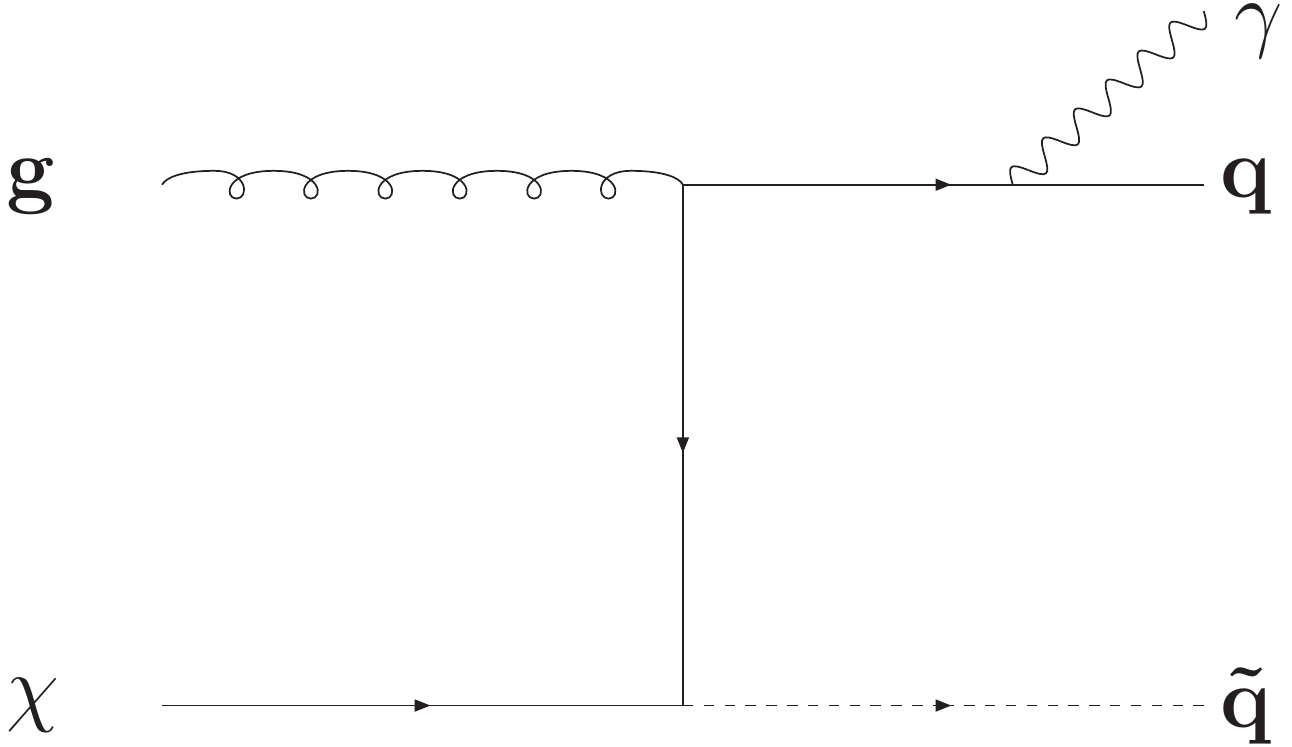}
\caption{A possible diagram with a gluon in the initial
  state. As explained in the text, we do not include $t$-channel diagrams like
  this in our calculation.} \label{fig:gluon}
\end{figure}

So far we have only considered electrons in the jet. Protons are most
likely present in the jet as well, and in some models that have gained some attention recently, they can even be the dominant component \cite{blobgeom}. In this section
we analyze the scattering of protons off of dark matter, with the
emission of photons in the final state. Because these photons are quite hard ($>1$ GeV), they will be emitted in the final state by the
quark {\em before} hadronization (notice that they can also be emitted from the squark
exchanged in the $s$-channel).  
The most straightforward approach to
compute this process is to study it at the parton level, neglecting details of the hadronization process. We will then have the same Feynman diagrams that
we already computed in previous sections, with the electrons replaced by quarks, and the
selectron replaced by a squark. If the incoming proton has momentum
$k$, the quark will carry momentum $x
k$, with $0<x<1$. The probability of having a quark carrying a given fraction of the proton's momentum is set by the proton's Parton Distribution Functions (PDF). The proton PDF indicate that, in general, gluons can carry a very large fraction of the proton momentum. One might
then worry that we should also consider Feynman diagrams with a gluon
in the initial state (see {\it e.g.} Fig.~\ref{fig:gluon}).
All such diagrams, however, can only be in the $t$-channel, so they can never
undergo a resonant enhancement. Therefore, their contribution to the
cross section is negligible compared to the $s$-channel diagrams that have resonant behaviors,
and we can neglect them for our current purposes.

The differential cross section, in the co-linear approximation, is then obtained from
the one already computed for electrons by replacing $k$ (the momentum of the
electron) with $xk$ (the momentum of the relevant quark, expressed as a fraction of the proton momentum), the mass of the selectron with
that of the squark, the mass of the electron in the log with a cutoff
of the order the QCD scale (we use here 100 MeV). The flux of photons is then given by
\beqn \label{eq:protonflux}
\frac{d\Phi_\gamma}{dE_\gamma}=\int dE \int_{x_{min}}^1 dx \sum_{i=u,d} f_i(x)\left(\frac{1}{M_\chi}
  \frac{d^2\sigma_{p+\chi \rightarrow \gamma + \dots}}{d\Omega
    dE_\gamma}\right)_{\cos \theta} \left(\frac{1}{d^2_{\rm
      AGN}}\frac{d\Phi^{\rm AGN}_{p}}{dE}\right) \delta_{\rm
  DM},
\eeqn
where $f_i(x)$ are the relevant PDF. (For our numerical calculations we used the
Mathematica package provided by the CTEQ collaboration \cite{CTEQ}.) 

As far as the AGN proton jet is concerned, we consider here two theoretically and observationally motivated distinct cases \cite{blobgeom}:
\begin{enumerate}
\item A jet consists of mono-energetic protons, which is believed to
  be the case for the innermost AGN jet regions \cite{blobgeom}. If the proton energy is $E_p$ we
  thus have
\beqn
\frac{d\Phi^{\rm AGN}_{p}}{dE}= k_1 \delta(E-E_p).
\eeqn
The expression above has to be normalized using the kinetic power of the protons in the jet, as deduced from
observations
\beqn
L_p = \int dE E \frac{d\Phi^{\rm AGN}_{p}}{dE},
\eeqn
which simply gives $k_1=L_p/E_p$.
\item The protons in the AGN jet feature a power law energy distribution. At some
  distance from the central compact source, the protons undergo a shock, which, assuming
  a simple Fermi acceleration mechanism, results in a protons spectrum proportional to $E_p^{-2}$.
In this case, one has
 \beqn
\frac{d\Phi^{\rm AGN}_{p}}{dE}= k_2 \left(\frac{m_p}{E}\right)^2.
\eeqn
Using the kinetic power to normalize the distribution, we finally get
\be
k_2=\frac{L_p}{m^2_p \ln \left(\frac{E_{max}}{E_{min}}\right)}.
\ee
\end{enumerate}
For Cen A we use a proton jet luminosity of $L_p = 3\times 10^{44}$ erg
s$^{-1}$, $E_{min}=10$ GeV, $E_{max}=10^7$ GeV, as determined from the interpretation of the Fermi-LAT observations reported in Ref.~\cite{fermicena}. The $L_p$ value is only 10 times larger than the kinetic
power we used for electrons. If we compare $k_2$ to $k_e$ we see that
the former is down by a factor of $m^2_e / m^2_p$ with respect to the
latter. This means that for protons we loose about six orders of
magnitude in the normalization of the energy distribution, compared to
electrons. We thus generically expect a much lower signal here than in the electron case. Recall that with the electrons the kinematic condition on the
minimum energy of the incoming electron read:
$E_{min}=\frac{E_\gamma}{1-\frac{E_\gamma}{M_\chi}(1-\cos\theta)}$. The
same condition applies to the incoming quark here, and it translates into a
condition on $x_{min}$, the lower limit of the integral in
Eq.~(\ref{eq:protonflux})
\beqn
x_{min}=\frac{E_\gamma / E_p}{1-\frac{E_\gamma}{M_\chi}(1-\cos\theta)}
\eeqn
Also, we must have $x_{min}<1$, that sets an upper limit on the photon
energy
\beqn
E_\gamma<\frac{1}{1/E_p+1/M_\chi (1-\cos \theta)}
\eeqn
Similarly to the electron case, when the minimum energy of the
incoming quark is such that we lose the first resonance ($s>M^2_{\tilde
  q}$), we expect the signal to drop. This happens for
\beqn
E_\gamma = \frac{M_\chi (M_{\tilde q}^2 -
  M_\chi^2)}{2M_\chi^2+(M_{\tilde q}^2 - M_\chi^2)(1-\cos\theta)}.
\eeqn

\begin{figure}[!t]
\centering
\includegraphics[width=180mm]{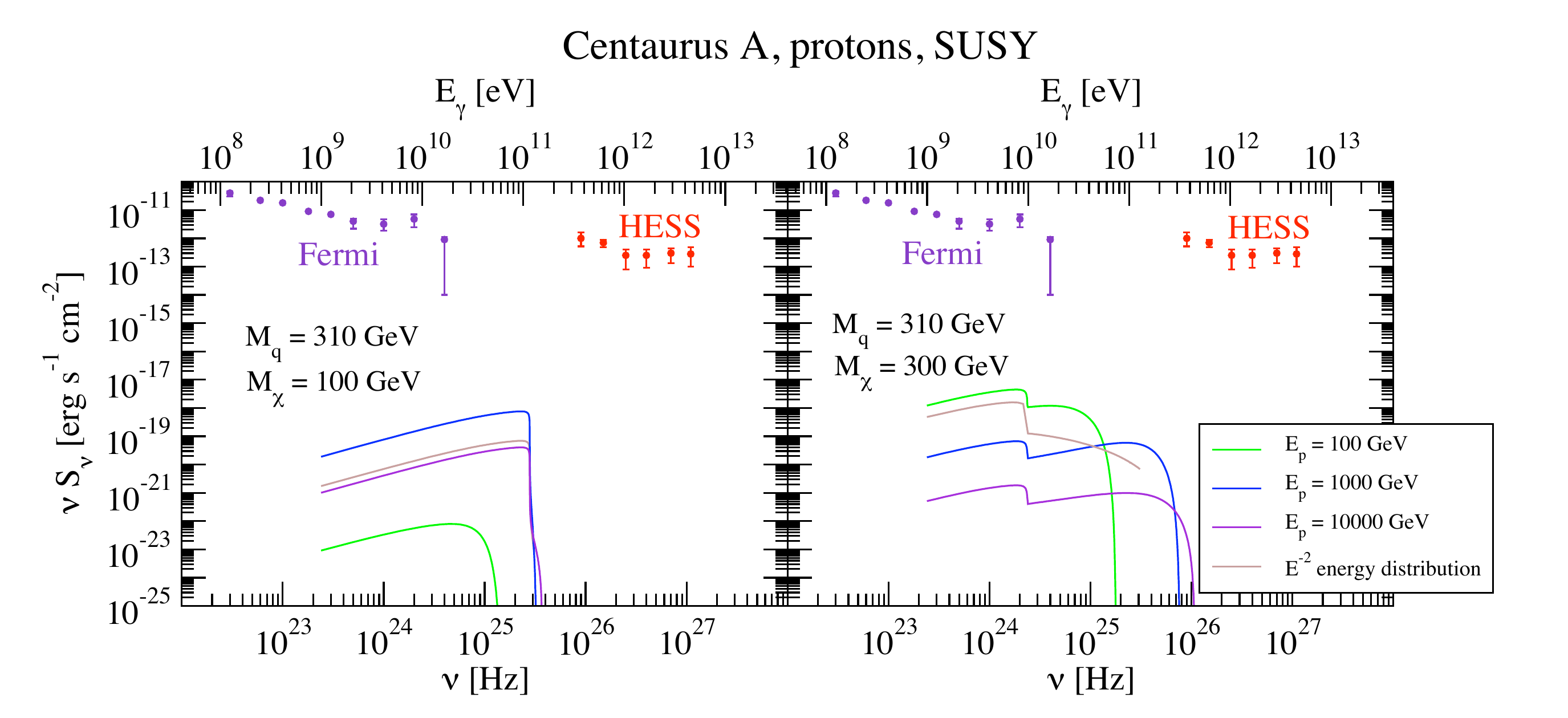}
\caption{The spectral energy distribution for photons resulting from proton-dark matter scattering in the Centaurus A AGN jet, under the assumption that the latter is dominantly composed of protons. In the Figure, the dark matter density normalization factor $\delta_{\rm DM}=10^8 M_\odot/pc^2$ and the proton jet luminosity $L_p=3\times 10^{44}$ erg/s$^{-1}$. The green, blue and purple lines correspond to mono-energetic proton jets with increasing energies set to $E_p=10^2,\ 10^3$ and $10^4$ GeV, while the brown line refers to an $E_p^2$ proton energy spectrum. The Fermi-LAT data are from Ref.~\cite{fermicena}, while the H.E.S.S. data from Ref.~\cite{Aharonian:2009xn}.}
\label{fig:protoni}
\end{figure}

We show in Fig~\ref{fig:protoni} our results for proton jets in the Cen A AGN, which confirm the theoretical expectations outlined above. In the Figure, we set $\delta_{\rm DM}=10^8 M_\odot/pc^2$ and $L_p=3\times 10^{44}$ erg/s$^{-1}$. The green, blue and purple lines correspond to mono-energetic proton jets with increasing energies set to $E_p=10^2,\ 10^3$ and $10^4$ GeV, while the brown line refers to an $E_p^2$ proton energy spectrum. Note that when the
masses of neutralino and of the squarks (which, similarly to the electron, we here assume to be all degenerate in mass) are close to each other (as in the right panel), the signal
drops by a relatively small amount, as it is still enhanced by the second
resonance, $s'=M^2_{\tilde q}$, that depends on $\omega$, until it
falls down again at even larger energies because the integral over $x$ is restricted to
regions of high $x$, where the PDF drops.
In the plot on the left, for $E_p=100$ GeV we never hit the first
resonance. This is the reason why the signal is lower than for all other assumptions on the proton spectrum, and why the signal drops down smoothly, simply due to the proton PDF.

The Figures suggest that in the case of proton-dominated jets, the detection of photons out of proton-dark matter scattering is always several orders of magnitude smaller than the gamma-ray output from nearby AGN jets as detected by Fermi-LAT and by H.E.S.S. The latitude one has in the overall normalization factors does not seem to allow for an enhancement as large as it would be needed to get a detectable signal. While in Fig.~\ref{fig:protoni} we consider only the case of Cen A, we also verified that the situation is even less promising for M87.

\section{Conclusions}\label{sec:conclusions}

Recent data from the Fermi Large Area Telescope and from H.E.S.S. on the flux of gamma rays from the local AGN in Centaurus A and M87 motivate the present critical re-assessment of the possibility of detecting the scattering of high-energy electrons in AGN jets off of dark matter. This novel indirect dark matter detection method was originally proposed in Ref.~\cite{Bloom:1997vm} where, however, for several reasons reviewed here, pessimistic conclusions were reached. In the present study we re-evaluated both the dark matter density distribution around compact objects in the light of the new theoretical results of \cite{Gondolo:1999ef} and \cite{Gnedin:2003rj} and of the observational results of \cite{Romanowsky:2000zb, Kraft:2003gp}; we modeled the AGN jets after the significant new observational results of Ref.~\cite{fermicena, fermim87}; finally, and more crucially, we carried out the complete calculation of the relevant electron (and proton) scattering cross section off of dark matter. 

We presented the full calculation for the case of supersymmetric neutralinos, and we presented an approximate, but, we argued, very accurate, result for the lightest Kaluza-Klein particle of Universal Extra Dimensions. The most important result that this calculation produced was the realization that the scattering process can be resonantly enhanced. Not only does this resonance produce much larger scattering rates than in the approximate dimensional analysis of \cite{Bloom:1997vm}, but it also dictates a very distinctive spectral feature, corresponding to the kinematic limit where the scattering can no longer occur resonantly. The spectral feature manifests in the final photon spectrum as a dramatic cutoff to an otherwise very hard spectrum. 

We argued that in view of the novel results of the present re-assessment of the dark matter search strategy outlined in \cite{Bloom:1997vm}, a large enough signal might be expected, especially for suitably light dark matter particle masses and for low enough masses associated to the dark sector partner of the electron (i.e. the selectron for the supersymmetric case and the Kaluza-Klein electron for UED). We compared our predictions with the actual gamma-ray data collected by Fermi-LAT, and found that for reasonable assumptions on the AGN jet energy content, spectrum and composition, a signal from electron-dark matter scattering at the level of the detected photon flux can be expected. We showed that photon self-absorption is irrelevant for the attenuation of the resulting gamma-ray flux, and that if the AGN jets are proton- instead of electron-dominated, the expected photon flux is likely much below the detectable intensity.

While the complicated nature of the gamma-ray emission from AGNs naturally prevents any conclusion on the possibility that the detected photons be exotic and possibly resulting from the interactions of high-energy electrons in the jet with the dense dark matter environment, it is exciting that such a signal might in principle be detectable. For Centaurus A, in particular, the difficulty, pointed out in Ref.~\cite{fermicena}, in explaining the low energy and the high energy emissions with a single-zone synchrotron/synchrotron self Compton (SSC) model leaves some room to explore other possibilities, such as the one proposed in this study. In the optimistic scenario that future collider results point towards a dark sector spectrum with features similar to those expected to produce a signal in AGN jets, further studies of the process examined here will most certainly deserve considerable attention.

\section*{Acknowledgements} \noindent SP would like to greatly thank Lars Bergstrom, who originally suggested to him to investigate the processes studied in this work, and who gave very useful feedback on this manuscript at various stages. This work is supported by NASA grant NNX08AV72G. SP is also partly supported by NSF grant PHY-0757911-001 and by an Outstanding Junior Investigator Award from the US Department of Energy.  We gratefully acknowledge conversations, suggestions and help from Elliott Bloom, Guido Festuccia, Howard Haber, Michael Peskin, Enrico Ramirez-Ruiz, Aaron Romanowski, G.~E.~Romero, David Rosario and Tim Tait.

\appendix

\section{Particle energy distributions in the AGN jet boosted frame} \label{app:special}
We consider a model for the AGN jet consisting of a blob moving along the $z$ axis in the black hole rest frame with a bulk Lorentz factor $\Gamma_B = (1-\beta_B^2)^{-1/2}$. The $z'$ axis of the blob frame is assumed to lie along the same direction as $z$, $\theta'$ and $\theta$ are the polar angles in the blob and in the black hole frame, respectively. We define $\mu' \equiv \cos \theta'$ and $\mu \equiv \cos \theta$. Primed quantities refer to the blob frame, unprimed quantities to the black hole frame.

\begin{figure}[!t]
\centering\label{fig:mumup}
\includegraphics[width=50 mm]{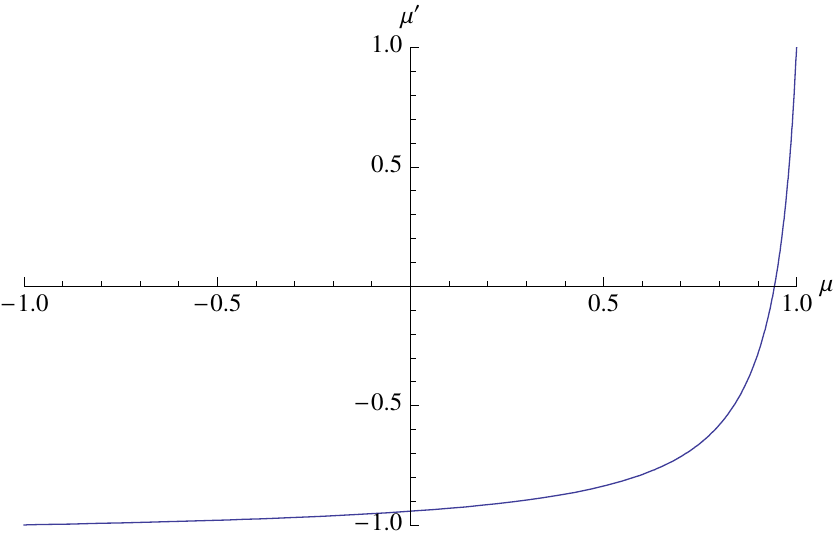}
\caption{The polar angle $\mu'=\cos \theta'$ in the blob frame versus the corresponding angle $\mu=\cos\theta$ in the black hole frame.}
\end{figure}

Consider electrons moving isotropically in the blob rest frame with
the following power law distribution
\be \label{eq:distr}
\frac{d\Phi_e^{\rm AGN}}{d\gamma'}(\gamma') d\gamma' d\mu' = \frac{1}{2} k_e \gamma'^{-s_1}\left[1+\left(\frac{\gamma'}{\gamma'_{\rm br}}\right)^{s_2-s_1}\right]^{-1} d\gamma' d\mu'.
\ee
Our goal is to boost this spectral distribution and calculate it as resulting in the black hole frame. The components of the electron velocity parallel and perpendicular to the $z$ axis transform, respectively, as
\be
\beta'_\parallel = \frac{\beta \cos \theta-\beta_B}{1-\beta_B \beta \cos \theta},\qquad \qquad \beta'_\perp = \frac{\beta \sin \theta}{\Gamma_B(1-\beta_B\beta \cos \theta)},
\ee
and it is straightforward to compute
\be
\gamma'=(1-\beta'^2_\parallel-\beta'^2_\perp)^{-1/2} = (1-\beta_B\beta \mu)\gamma\Gamma_B.
\ee
Since the electrons are highly relativistic, we can safely approximate $\beta \simeq 1$ in the above and write
\be
\gamma'=(1-\beta_B \mu)\gamma\Gamma_B.
\ee
With this approximation, the polar angle transforms as
\be
\mu' = \frac{\mu-\beta_B}{1-\beta_B \mu}.
\ee
Working out the differentials for the two Equations above is easy, and we can then simply substitute in Eq.~(\ref{eq:distr}) to find
\be
\frac{d\Phi_e^{\rm AGN}}{d\gamma'}(\gamma') d\gamma' d\mu' = \frac{d\Phi_e^{\rm AGN}}{d\gamma} (\gamma\Gamma_B(1-\beta_B\mu)) \ (\Gamma_B(1-\beta_B\mu)d\gamma) \frac{d\mu}{\Gamma_B^2(1-\beta_B\mu)^2}.
\ee 
It is instructive to look at the plot of $\mu'$ as a function of $\mu$, which we show in Fig.~\ref{fig:mumup} for $\Gamma_B=3$, to appreciate how most of the electrons are seen to move close to the forward direction.

\section{The differential cross section for the supersymmetric case} \label{sec:fullcross}
In this appendix we compute the differential cross section
$\frac{d^2\sigma}{dE_\gamma d\Omega}$ for the process
``electron + neutralino $\rightarrow$ electron + neutralino +
photon'' in the context of the supersymmetric extension of the Standard Model, and for degenerate selectron masses. As discussed above, when the exchanged
selectron is on-shell, the only relevant Feynman diagrams are the ones
shown in Fig.~\ref{fig:SUSYdiagrams}.
\begin{enumerate}
\item Bremsstrahlung form the initial electron:
\beqn
{\cal M}_1^\mu&=&ie\frac{1}{t}\Pi_{s'}
\bar e'(a_LP_R+a_R P_L)\chi'\,\bar\chi(\sk-\sq)\gamma^\mu (a_LP_L+a_RP_R)e,
\eeqn
where $a_L$ ($a_R$) is the coupling for the vertex left (right)-handed
electron - selectron - neutralino\footnote{$a^2_L=2g^2\tan^2 \theta_W$
  and $a^2_R=\frac{1}{4}a^2_L$, with $g$ the Standard Model $SU(2)$
  gauge coupling and $\theta_W$ the Weinberg angle. For the derivation of these couplings in the MSSM see, {\it e.g.}, the appendix of Ref.~\cite{Edsjo:1997hp}.}, 
and $P_{L,R}=\frac{1}{2}(1\mp\gamma_5)$ are the corresponding projectors. Also, the identity 
$(\bar{e}_L)=\bar eP_R$ was used.
\item Bremsstrahlung form the final electron:
\beqn
{\cal M}_2^\mu&=&ie\frac{1}{t'}\Pi_{s}
\bar e'(a_LP_R+a_R P_L)\gamma^\mu(\sk'+\sq)\chi'\,\bar\chi(a_LP_L+a_RP_R)e.
\eeqn
\item Photon emission from the selectron:
\beqn
{\cal M}_3^\mu&=&ie\Pi_{s}\Pi_{s'}(p+k+p'+k')^\mu
\left(1-i\frac{\Gamma}{\sqrt{s}+\sqrt{s'}}\right)
\bar e'(a_LP_R+a_R P_L)\chi'\,\bar\chi(a_LP_L+a_RP_R)e.
\eeqn
\end{enumerate}

It can be shown that the full amplitude is gauge invariant,
$\sum_{i=1,2,3}q_\mu{\cal M}_i^\mu=0$. 
To this end, one needs the definitions given above, and the identity 
$q_\mu (p+k)^\mu=q_\mu (p'+k')^\mu=\frac{s-s'}{2}$. Note that the Feynman rule 
in the last amplitude is a generalized one for unstable particles, that was 
introduced to make the amplitude explicitely gauge 
invariant. It reduces to the usual one for $\Gamma=0$. To calculate the spin-averaged squared amplitude
 we first consider terms with a $1/t'$ divergence, which will give
the leading logarithmic terms. These are

\beqn
&&\frac{1}{4}\sum_{\lambda,spins}|\varepsilon_\mu^*(\lambda){\cal M}_2^\mu|^2
=2e^2(a_L^4+a_R^4)\frac{|\Pi_s|^2}{t'}(pk)(p'q), \nn\\
&&\frac{1}{4}\sum_{\lambda,spins}
\left\{\varepsilon_\mu^*(\lambda){\cal M}_2^\mu 
\varepsilon_\nu(\lambda){\cal M}_3^{*\nu}+h.c\right\}=-2e^2(a_L^4+a_R^4)\frac{|\Pi_s|^2}{t'}|\Pi_{s'}|^2
[s'-M_{\tilde e}^2+\frac{\sqrt{s'}\Gamma^2}{\sqrt{s}+\sqrt{s'}}]
(p'k')4(pk)^2,\nn\\
&&\frac{1}{4}\sum_{\lambda,spins}
\left\{\varepsilon_\mu^*(\lambda){\cal M}_2^\mu 
\varepsilon_\nu(\lambda){\cal M}_1^{*\nu}+h.c\right\}\nn\\
&&=2e^2(a_L^4+a_R^4)\frac{1}{t'}{\rm Re}(\Pi_{s}\Pi_{s'}^*)
\left[(pk)(p'k')+(p'k')(pk')+(pk)(kk')
-\frac{(kk')}{t}[4(p'k')(pk)+(pq)(s-s')]\right],
\eeqn
where we used $t'=(k'+q)^2=2(k'q)$ and we neglected $(k'q)$ in the
numerator where possible. For future convenience we can group them into 
\be
|{\cal M}|^2_{\rm log}=\frac{1}{4}\sum_{\lambda,spins}|\varepsilon_\mu^*(\lambda){\cal M}_2^\mu|^2+\frac{1}{4}\sum_{\lambda,spins}\left\{\varepsilon_\mu^*(\lambda){\cal M}_2^\mu 
\varepsilon_\nu(\lambda){\cal M}_3^{*\nu}+h.c\right\}+\frac{1}{4}\sum_{\lambda,spins}
\left\{\varepsilon_\mu^*(\lambda){\cal M}_2^\mu 
\varepsilon_\nu(\lambda){\cal M}_1^{*\nu}+h.c\right\}.
\ee
 The remaining 
pieces that do not have large logarithmic enhancements are
\beqn
\frac{1}{4}\sum_{\lambda,spins}|\varepsilon_\mu^*(\lambda){\cal M}_1^\mu|^2
&=&2e^2(a_L^4+a_R^4)\frac{|\Pi_s|^2}{|t|}(p'k')(pq),\nn\\
\frac{1}{4}\sum_{\lambda,spins}|\varepsilon_\mu^*(\lambda){\cal M}_3^\mu|^2
&=&-2e^2(a_L^4+a_R^4)|\Pi_s|^2|\Pi_{s'}|^2(pk)(p'k')(s+s')
\left[1+\frac{\Gamma^2}{(\sqrt{s}+\sqrt{s'})^2}\right],\nn\\
\frac{1}{4}\sum_{\lambda,spins}
\left\{\varepsilon_\mu^*(\lambda){\cal M}_2^\mu 
\varepsilon_\nu(\lambda){\cal M}_3^{*\nu}+h.c\right\}
&=&-2e^2(a_L^4+a_R^4)\frac{|\Pi_s|^2}{t}|\Pi_{s'}|^2
[s-M_{\tilde e}^2+\frac{\sqrt{s}\Gamma^2}{\sqrt{s}+\sqrt{s'}}]
(p'k')[4(pk)(p'k')-\frac{st}{2}].\nn\\
\eeqn
Again, we group them into the following term:
\be
|{\cal M}|^2_{\rm no log}=\frac{1}{4}\sum_{\lambda,spins}|\varepsilon_\mu^*(\lambda){\cal M}_1^\mu|^2+\frac{1}{4}\sum_{\lambda,spins}|\varepsilon_\mu^*(\lambda){\cal M}_3^\mu|^2+\frac{1}{4}\sum_{\lambda,spins}\left\{\varepsilon_\mu^*(\lambda){\cal M}_2^\mu 
\varepsilon_\nu(\lambda){\cal M}_3^{*\nu}+h.c\right\}.
\ee
The differential cross section is then
\be
d\sigma=(2\pi)^4\delta^4(p+k-p'-k'-q)\frac{1}{4M_\chi E}
\frac{d^3 k'}{2E'(2\pi)^3}\frac{d^3 q}{2E_\gamma(2\pi)^3}
\frac{d^3 p'}{2E'_N(2\pi)^3}(|{\cal M}|^2_{\rm log}+|{\cal M}|^2_{\rm
  no log}).
\ee
The kinematics were discussed in section~\ref{sec:cross}. Note that,
because the scalar products
\beqn
(p'k')&=&(p+k-k'-q,k')\to(pk)-(p+k,q)\nn\\
(p'q)&=&(p+k-k'-q,q)\to(p+k,q)
\eeqn
do not depend on $k'$, the integral over the solid $k'$ angle is
trivial for most terms in the squared amplitude. The final result for the cross section in the supersymmetric case is
\beqn \label{eq:fullcross}
\frac{d^2\sigma}{dE_\gamma
  d\Omega}&=&\frac{1}{(2\pi)^5}\frac{1}{32M_\chi EE'_N}\left(|{\cal
    M}|^2_{\rm log}t'\int d\Omega_{k'} \frac{E_\gamma E'}{t'} + 4\pi |{\cal M}|^2_{\rm
  no log}\right)\nn\\
&=&\frac{\pi}{(2\pi)^5}\frac{1}{32M_\chi EE'_N}\left(|{\cal
    M}|^2_{\rm log}t'\ln\left(\frac{4E'^2}{m_e^2}\right) + 4 |{\cal M}|^2_{\rm
  no log}\right).
\eeqn

\end{document}